\documentclass[a4paper,11pt]{article}
\pdfoutput=1
\usepackage{jheppub} 
\usepackage{graphicx,color,rotating}
\usepackage{hyperref}
\usepackage{epsfig,color}
\usepackage{slashed}
\usepackage{amsfonts}
\usepackage{ulem}

\newcommand{\gsim}{\raisebox{-0.13cm}{~\shortstack{$>$ \\[-0.07cm] $\sim$}}~}

\usepackage{tabularx}
\usepackage{graphicx}
\usepackage{adjustbox}
\usepackage{tabularx}

\begin{document}
\title{Axion-like particles, two-Higgs-doublet models, leptoquarks, and the electron and muon $g-2$}

\renewcommand{\thefootnote}{\arabic{footnote}}

\author{
Wai-Yee Keung$^{1}$,  Danny Marfatia$^{2}$, and
Po-Yan Tseng$^{3}$}
\affiliation{
$^1$ Department of Physics, University of Illinois at Chicago,
Illinois 60607 USA \\
$^2$ Department of Physics \& Astronomy, University of Hawaii at Manoa,
Honolulu, HI 96822, USA \\
$^3$ Department of Physics and IPAP, Yonsei University,
Seoul 03722, Republic of Korea \\
}
\date{\today}

\abstract{
Data from the Muon g-2 experiment and measurements of the fine structure constant suggest that the anomalous magnetic moments of the muon and electron are at odds with standard model expectations. We survey the ability of axion-like particles, two-Higgs-doublet models and leptoquarks to explain the discrepancies. We find that accounting for other constraints, all scenarios except the Type-I, Type-II and Type-Y two-Higgs-doublet models fit the data well.
}

\maketitle

\section{Introduction}

The high-intensity and high-precision frontiers are ideal for the
search for new physics that couples very feebly with the standard model (SM) sector.
%
A long-standing and perhaps best known example that indicates such physics is
the $3.7\sigma$ anomaly in the anomalous magnetic moment of the muon $a_\mu=(g-2)_\mu/2$:
\begin{eqnarray}
\Delta a^{\rm BNL}_\mu &=& a^{\rm BNL}_\mu-a^{\rm SM}_\mu = (279\pm 76) \times 10^{-11}\,,
\end{eqnarray}
where $a^{\rm BNL}_\mu=(116592089 \pm 63)\times 10^{-11}$~\cite{Bennett:2006fi,Zyla:2020zbs} 
and the SM expectation is $a^{\rm SM}_\mu=(116591810\pm 43)\times 10^{-11}$~\cite{Fermilab_muon}.
A new lattice QCD calculation of the hadronic vacuum polarization suggests that the BNL measurement is compatible with
the SM and that no new physics need be incorporated~\cite{Borsanyi:2020mff}. 
Until this result is confirmed, we subscribe to the SM value of 
Ref.~\cite{Fermilab_muon}.
Recently, the Muon g-2 experiment at Fermilab reported the value, 
$a^{\rm FNAL}_\mu=(116592040 \pm 54)\times 10^{-11}$~\cite{Fermilab_muong2}, i.e, 
\begin{eqnarray}
\Delta a_\mu^{\rm FNAL} &=& a^{\rm FNAL}_\mu-a^{\rm SM}_\mu = 
(230 \pm 69)\times 10^{-11}\,,
\end{eqnarray} 
which is a 3.3$\sigma$ discrepancy.
The combined significance of the anomaly from the Fermilab and BNL measurements is 4.25$\sigma$ with~\cite{Fermilab_muong2}
\begin{eqnarray}
\Delta a_\mu &=& a^{\rm exp}_\mu-a^{\rm SM}_\mu = (251 \pm 59)\times 10^{-11}\,.
\end{eqnarray}

Interestingly, new precise measurements of the fine-structure constant $\alpha$ imply a discrepancy in the anomalous magnetic moment of the
electron as well. A measurement of $\alpha$ at Laboratoire Kastler Brossel (LKB) with $^{87}$Rb atoms~\cite{Morel:2020dww}
improves the accuracy by a factor of 2.5 compared to 
the previous best measurement with $^{137}$Cs atoms at Berkeley~\cite{Parker:2018vye}. 
The LKB measurement deviates by $5.4\sigma$ from the Berkeley result.
With these two measurements of $\alpha$, 
the SM predictions for the electron anomalous magnetic moments, $a^{\rm LKB}_e$
and $a^{\rm B}_e$~\cite{Aoyama:2012wj,Aoyama:2019ryr},
 differ from the experimental measurement $a^{\rm exp}_e$~\cite{Hanneke:2008tm} at $1.6\sigma$ and $2.4\sigma$, respectively:
\begin{eqnarray}
\Delta a^{\rm LKB}_e &=& a^{\rm exp}_e-a^{\rm LKB}_e = (4.8\pm 3.0) \times 10^{-13}\,, \nonumber \\
\Delta a^{\rm B}_e &=& a^{\rm exp}_e-a^{\rm B}_e = (-8.8\pm 3.6) \times 10^{-13}\,. 
\end{eqnarray}
Note the opposite signs of $\Delta a^{\rm LKB}_e$ and $\Delta a^{\rm B}_e$.  

In this work, we study how well pseudoscalar axion-like particles (ALPs),
two-Higgs-doublet models (2HDMs), and leptoquarks (LQs) can provide
a common explanation of the anomalies in 
 $\Delta a_\mu$ and $\Delta a^{\rm LKB,\,B}_e$.

\section{Axion-like particles}

%

An ALP, in general, can couple to the photon and leptons via the
effective interactions~\cite{Marciano:2016yhf},
\begin{eqnarray}
\label{eq:ALP_interaction}
\mathcal{L}\supset \frac{1}{4}g_{a\gamma\gamma}a F_{\mu\nu}\tilde{F}^{\mu\nu}+iy_{a\ell}a \bar{\ell}\gamma_5 \ell\,,
\end{eqnarray}
where $g_{a\gamma \gamma}$ is a dimensionful coupling, and
$F_{\mu\nu}$ and $\tilde{F}^{\mu\nu}$ are the electromagnetic tensor 
and its dual, respectively. We can take $g_{a\gamma\gamma}$ to be positive by absorbing a phase into the definition of the field
$a$. Then the sign of $y_{a\ell}$ becomes physical.
If $\Lambda$ is the ultraviolet cut-off of the effective theory,
$g_{a\gamma\gamma}=2\sqrt{2}\alpha c_{a\gamma \gamma}/ \Lambda$ 
with dimensionless coupling $c_{a\gamma\gamma}$. 
The first term in Eq.~(\ref{eq:ALP_interaction}) induces the two loop light-by-light (LbL) diagram which is analogous to the
SM hadronic contribution from $\pi^0$ exchange~\cite{Jegerlehner:2009ry,Prades:2009aq,Dorokhov:2015psa}.
Both terms in Eq.(\ref{eq:ALP_interaction}) contribute 
to $g-2$ via Barr-Zee (BZ) diagrams~\cite{BZ}.
By only keeping the leading log, these contributions to $a_\ell$ give
\begin{eqnarray}
a_{\ell,a} & = & a^{\rm 1-loop}_{\ell,a}+a^{\rm BZ}_{\ell,a}+a^{\rm LbL}_{\ell,a}\,, \ \ \ \ {\text{where}}\\
a^{\rm 1-loop}_{\ell,a} & \simeq &
\frac{y^2_{a\ell}}{8\pi^2}\left( \frac{m_\ell}{m_a} \right)^2 f_A(m^2_\ell/m^2_a)
~\text{\cite{Chun:2015xfx}}
 \,, \nonumber \\
a^{\rm BZ}_{\ell,a} & \simeq & 
 \frac{m_\ell}{4\pi^2}
g_{a\gamma\gamma}y_{a\ell} \ln \frac{\Lambda}{m_a}~\text{\cite{Marciano:2016yhf}} \,, \nonumber \\
a^{\rm LbL}_{\ell,a} & \simeq & 
3 \frac{\alpha}{\pi} \left(\frac{m_\ell g_{a\gamma\gamma}}{4 \pi} \right)^2
\left(\ln \frac{\Lambda}{m_a} \right)^2~\text{\cite{Marciano:2016yhf}} \,.
\nonumber 
\end{eqnarray}
Here, $m_a$ is the ALP mass and the one-loop function for a pseudoscalar is~\cite{Chun:2015xfx}
\begin{eqnarray}
\label{eq:A_loop_function}
f_A(r)&=& \int^1_0 dx \frac{-x^3}{1-x+rx^2}\,.
\end{eqnarray}

ALP masses between 0.1 GeV to 10 GeV are allowed for non-negligible $g_{a\gamma\gamma}$~\cite{Jaeckel:2015jla}.
In particular, $m_a\leq 0.1$ GeV is restricted by beam dump experiments,
and LEP data on the decay $Z \to 3 \gamma$ 
 constrains $m_a\geq 10$ GeV via the process 
$e^+e^- \to \gamma^* \to a \gamma\to 3 \gamma$~\cite{Mimasu:2014nea}.
Also, $Z \to 2\gamma$ data at LEP provide a constraint if photons from $a \to 2\gamma$ are collimated
as a single photon. An upper bound $g_{a\gamma \gamma}\simeq \mathcal{O}(10^{-2})~{\rm GeV^{-1}}$ is obtained
for $1~{\rm MeV} \leq m_a\leq 10~{\rm GeV}$~\cite{Jaeckel:2015jla}.
For this coupling, unitarity  requires an upper bound, $\Lambda \simeq 1$~TeV~\cite{Marciano:2016yhf}.

To obtain parameter values preferred by the data, we separately fit $\Delta a_\mu$, $\Delta a^{\rm LKB}_e$, and $\Delta a^{\rm B}_e$, and also fit  the combinations, $\Delta a_\mu$ and $\Delta a^{\rm LKB}_e$,  and
 $\Delta a_\mu$ and $\Delta a^{\rm B}_e$. We do not fit $\Delta a^{\rm LKB}_e$ and $\Delta a^{\rm B}_e$ simultaneously. We use the following $\chi^2$ definitions:
   \begin{eqnarray}
\chi^2_{a_\mu} &\equiv & \frac{(a_{\mu,a}-\Delta a_\mu)^2}{(\sigma_{\Delta a_\mu})^2}\,,~~
\chi^2_{a^{\rm LKB}_e} \equiv  \frac{(a_{e,a}-\Delta a^{\rm LKB}_e)^2}{(\sigma_{\Delta a^{\rm LKB}_e})^2}\,,~~
\chi^2_{a^{\rm B}_e} \equiv \frac{(a_{e,a}-\Delta a^{\rm B}_e)^2}{(\sigma_{\Delta a^{\rm B}_e})^2}\,, \nonumber \\
\chi^2_{\rm LKB} &\equiv &\chi^2_{a_\mu}+\chi^2_{a^{\rm LKB}_e}
\,,~~
\chi^2_{\rm B} \equiv \chi^2_{a_\mu}+\chi^2_{a^{\rm B}_e}
\,.
\end{eqnarray}
Similar definitions will apply for 2HDMs and leptoquarks. 

\begin{figure}[t]
\centering
\includegraphics[height=1.5in,angle=0]{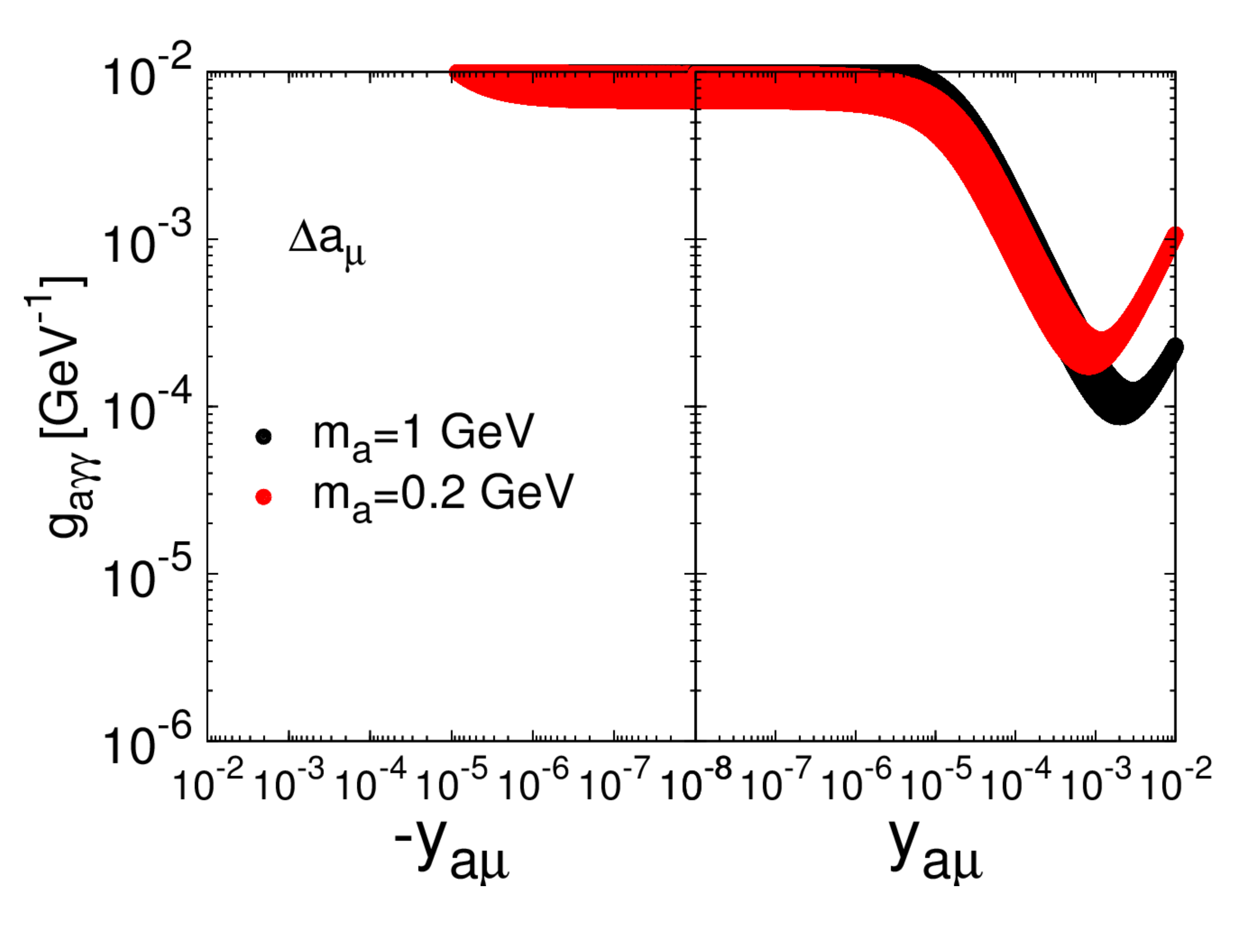}
\includegraphics[height=1.5in,angle=0]{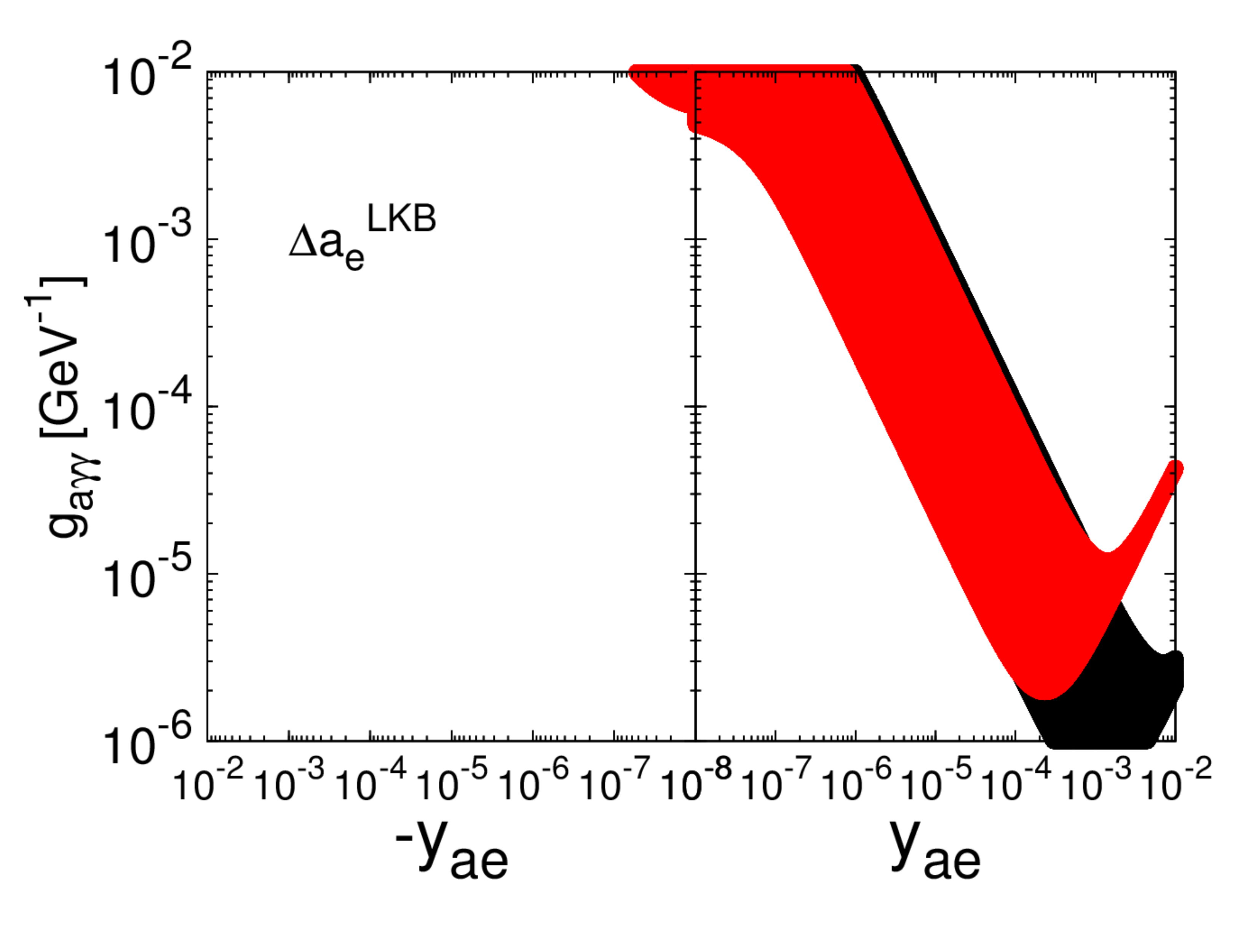}
\includegraphics[height=1.5in,angle=0]{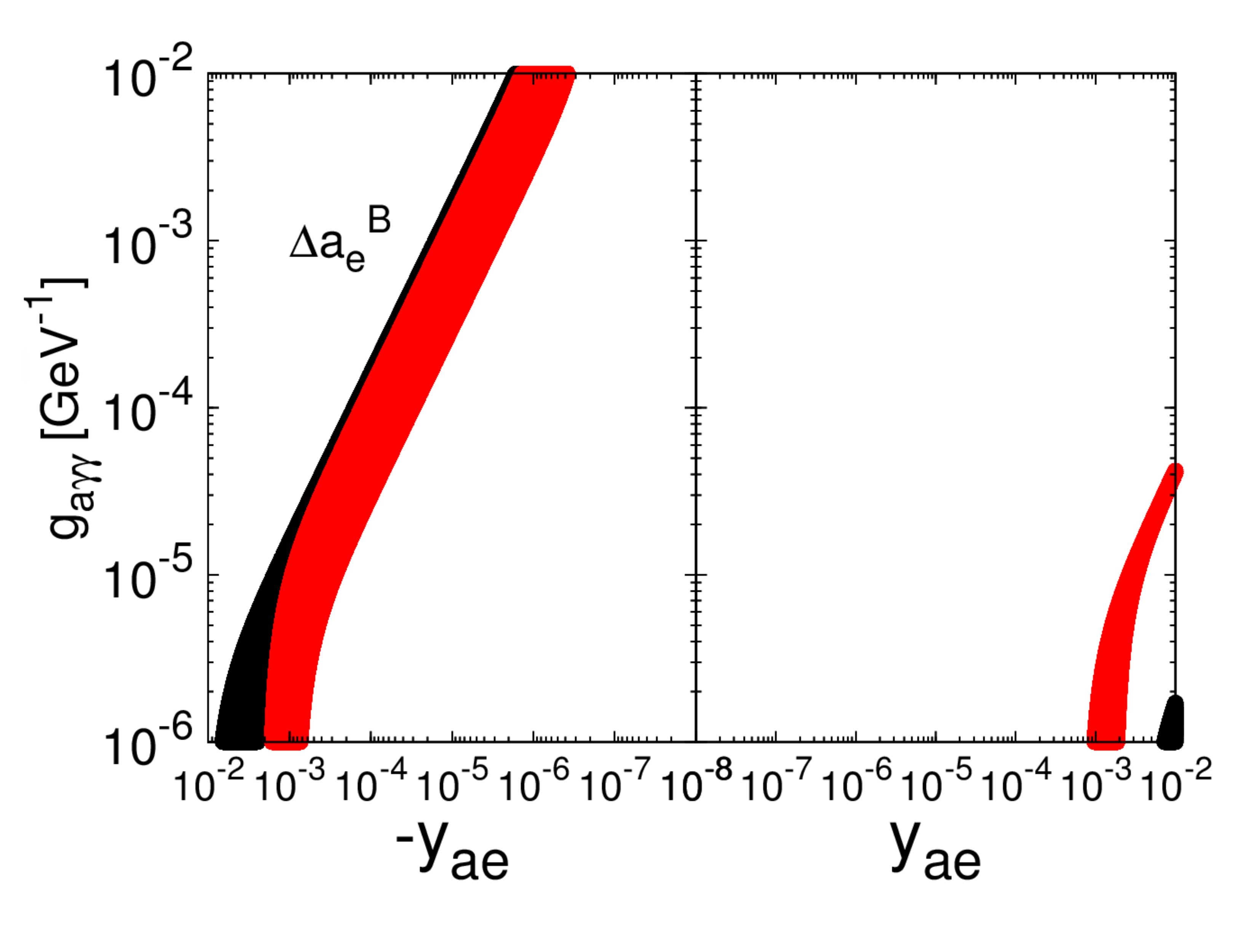}
\caption{\small \label{fig:scan1_g2_amu}
{\bf ALP-1:} The 1$\sigma$ regions preferred by $\Delta a_\mu$, $\Delta a^{\rm LKB}_e$, and $\Delta a^{\rm B}_e$ for $m_a=1$~GeV (black)
and $m_a=0.2$~GeV (red). 
}
\end{figure}

\begin{figure}[t]
\centering
\includegraphics[height=1.75in,angle=0]{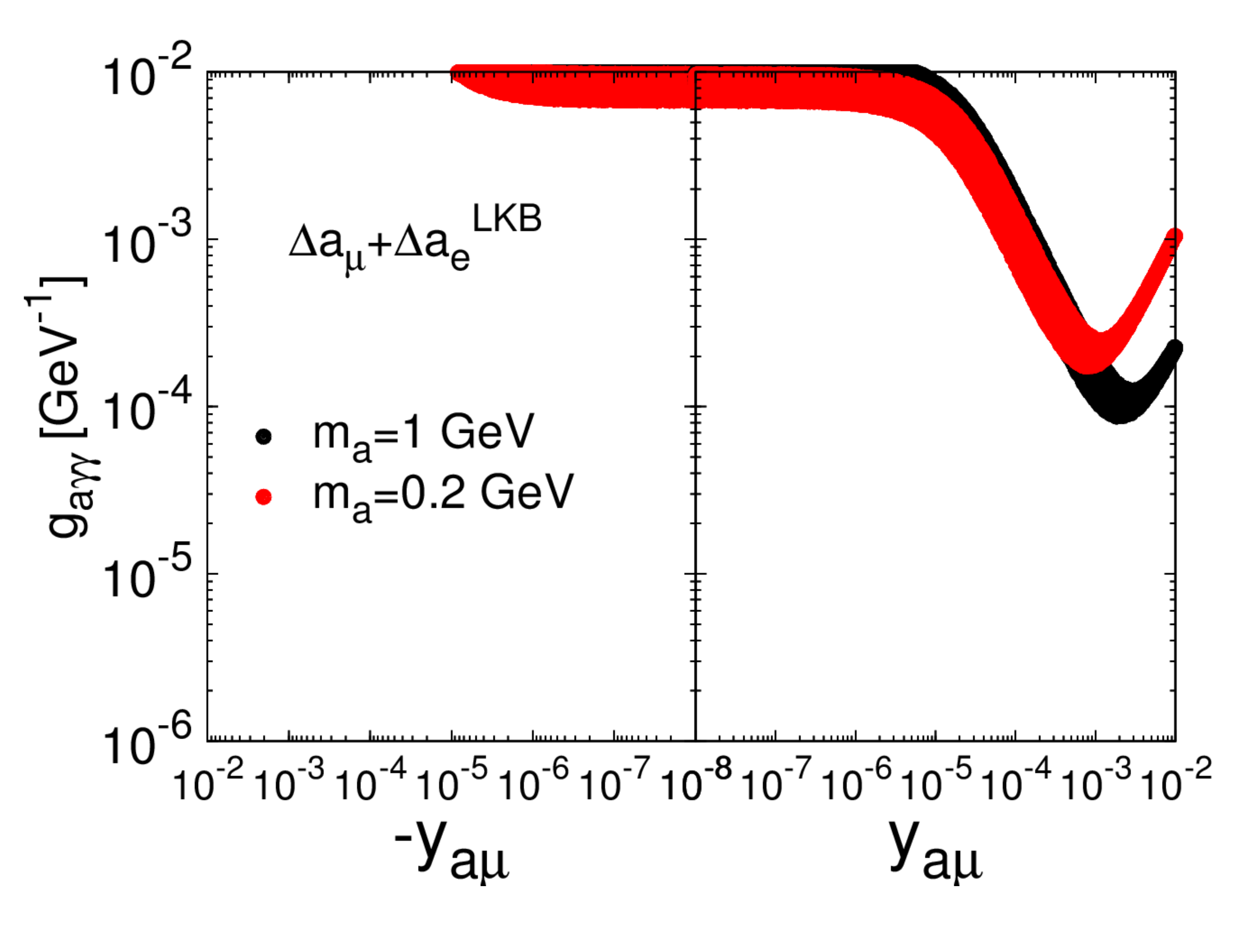}
\includegraphics[height=1.75in,angle=0]{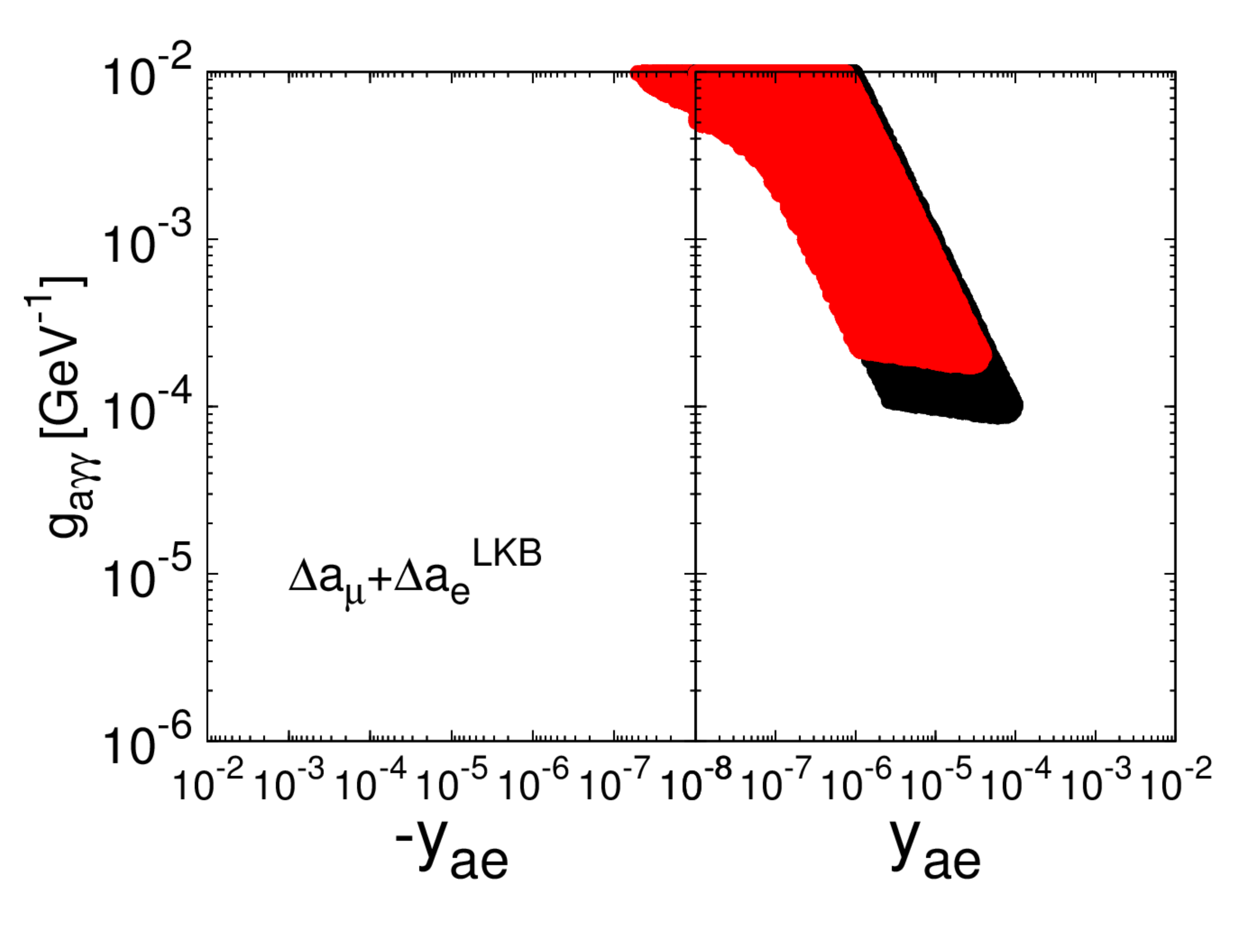}
\includegraphics[height=1.75in,angle=0]{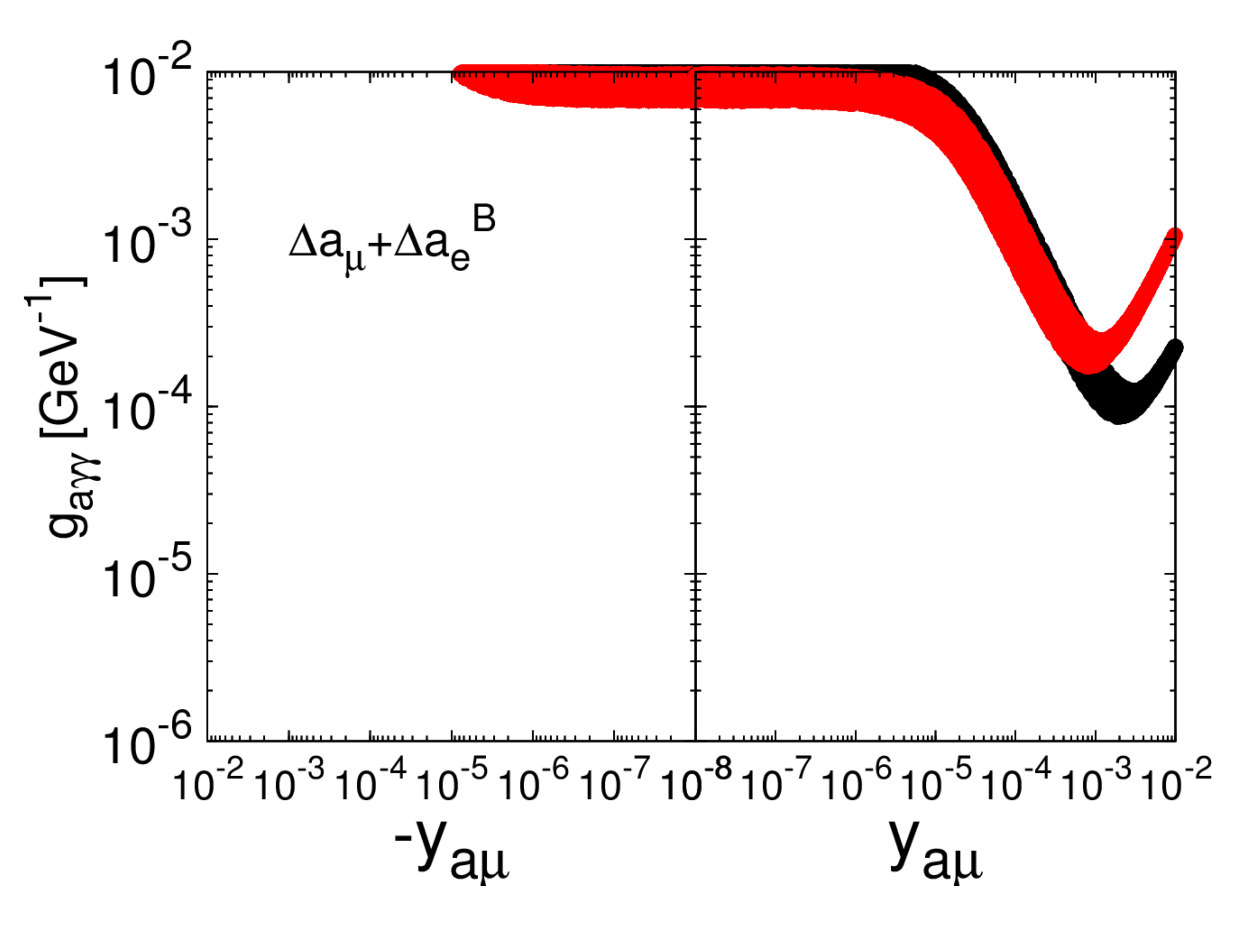}
\includegraphics[height=1.75in,angle=0]{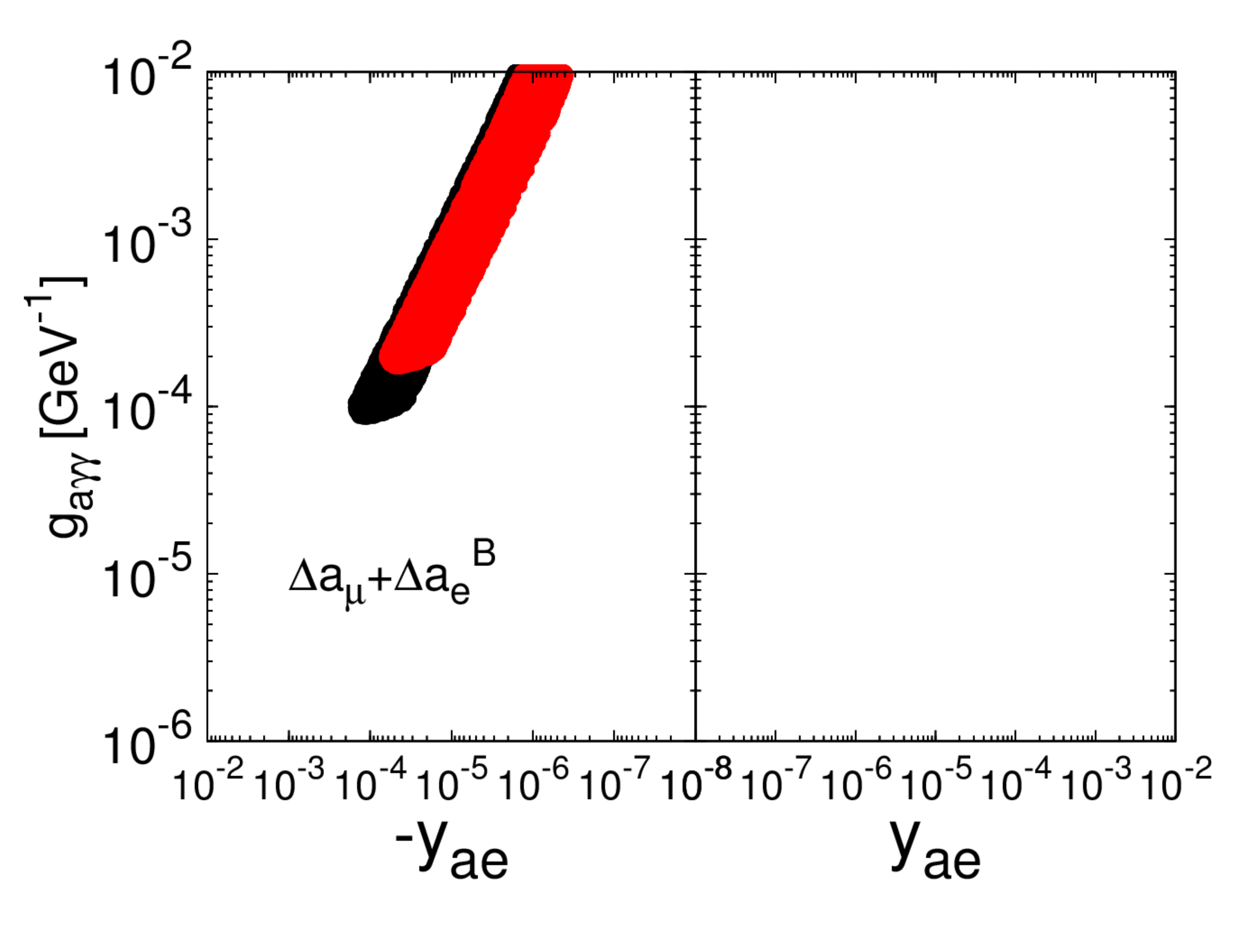}
\caption{\small \label{fig:scan1_g2_amu_ae}
{\bf ALP-1:} The 1$\sigma$ allowed regions from a combined fit to $\Delta a_\mu$ and $\Delta a^{\rm LKB}_e$ (upper panels) and
 $\Delta a_\mu$ and $\Delta a^{\rm B}_e$ (lower panels), for $m_a=1$~GeV (black)
and $m_a=0.2$~GeV (red). 
}
\end{figure}

\begin{figure}[t]
\centering
\includegraphics[height=1.5in,angle=0]{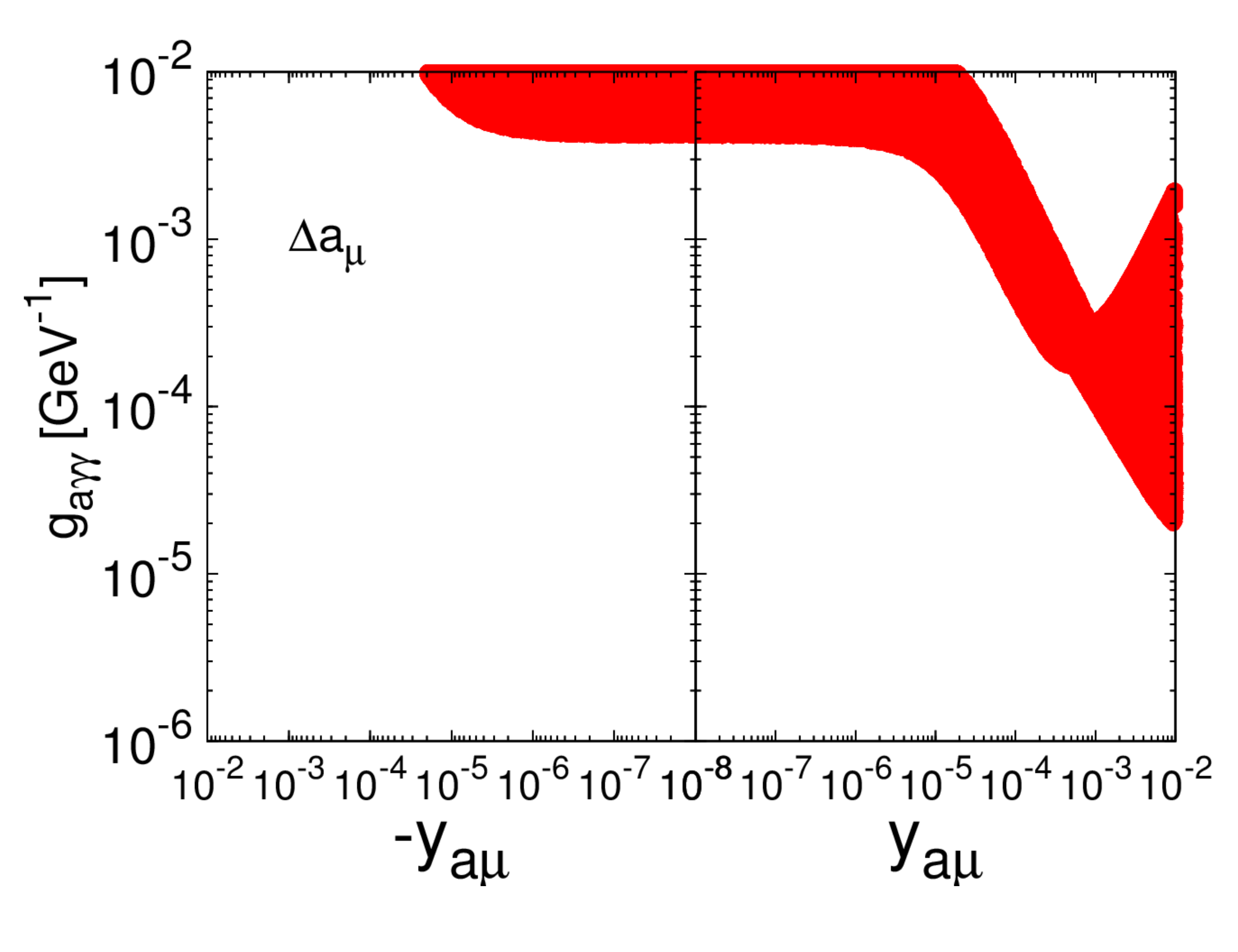}
\includegraphics[height=1.5in,angle=0]{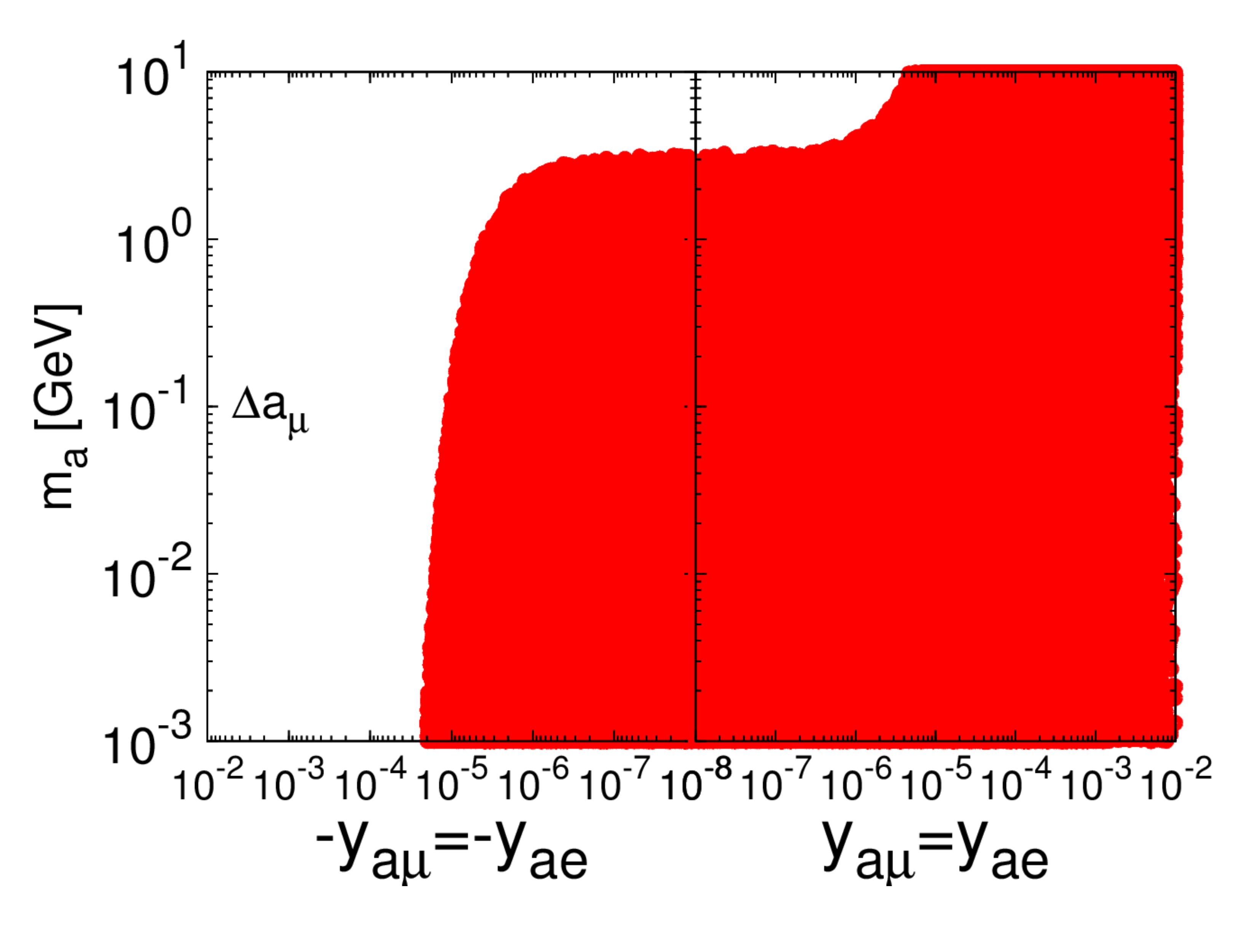}
\includegraphics[height=1.5in,angle=0]{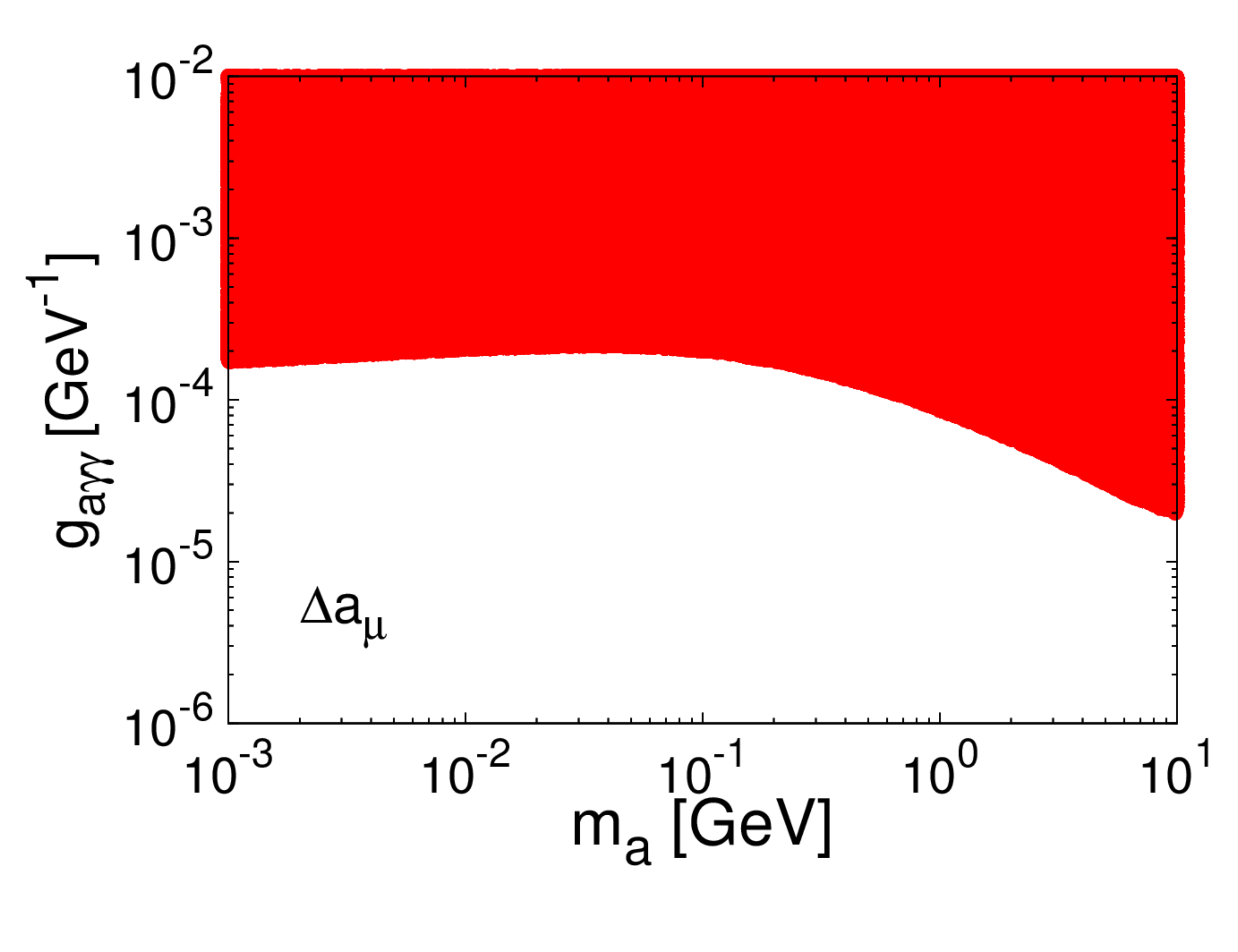}
\includegraphics[height=1.5in,angle=0]{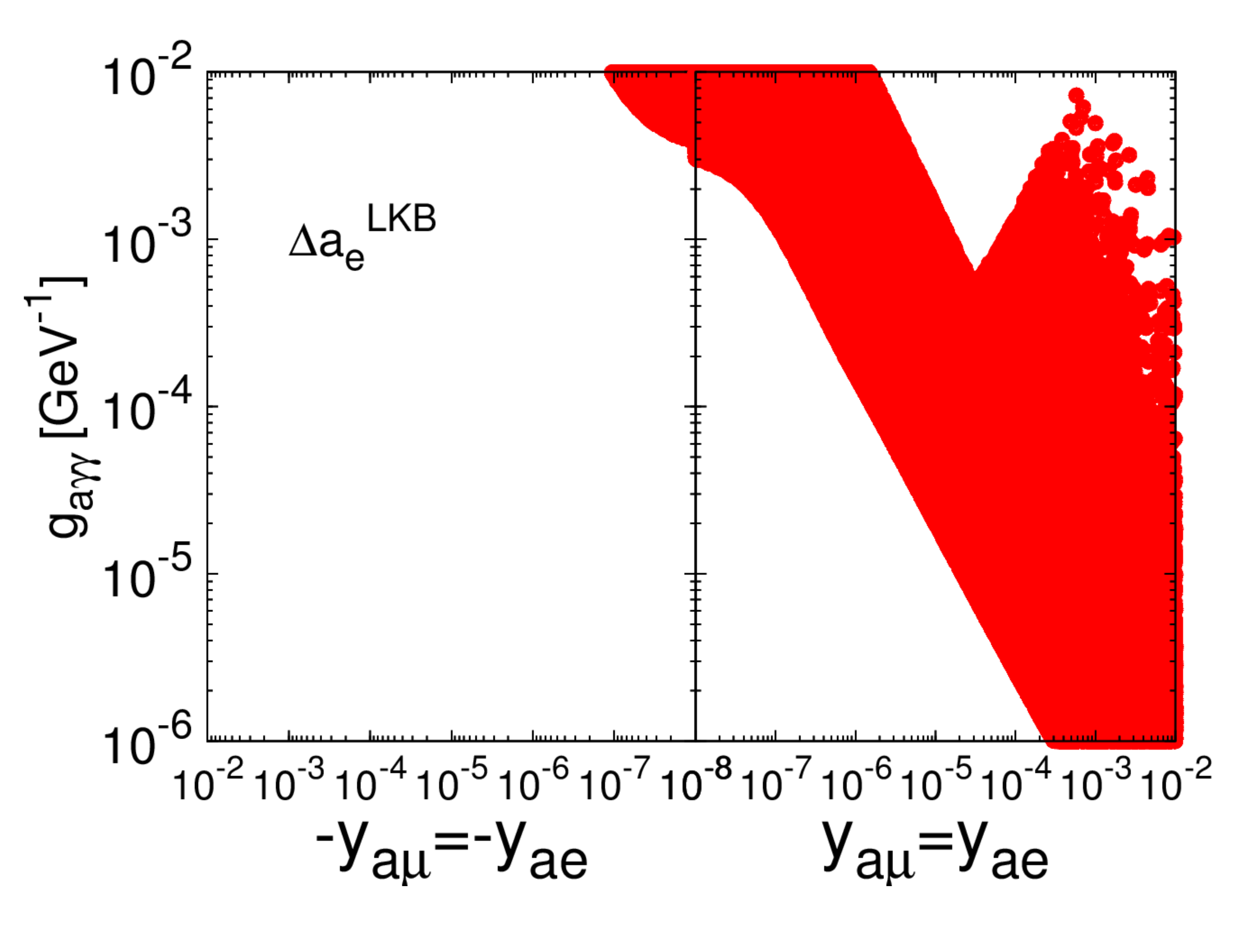}
\includegraphics[height=1.5in,angle=0]{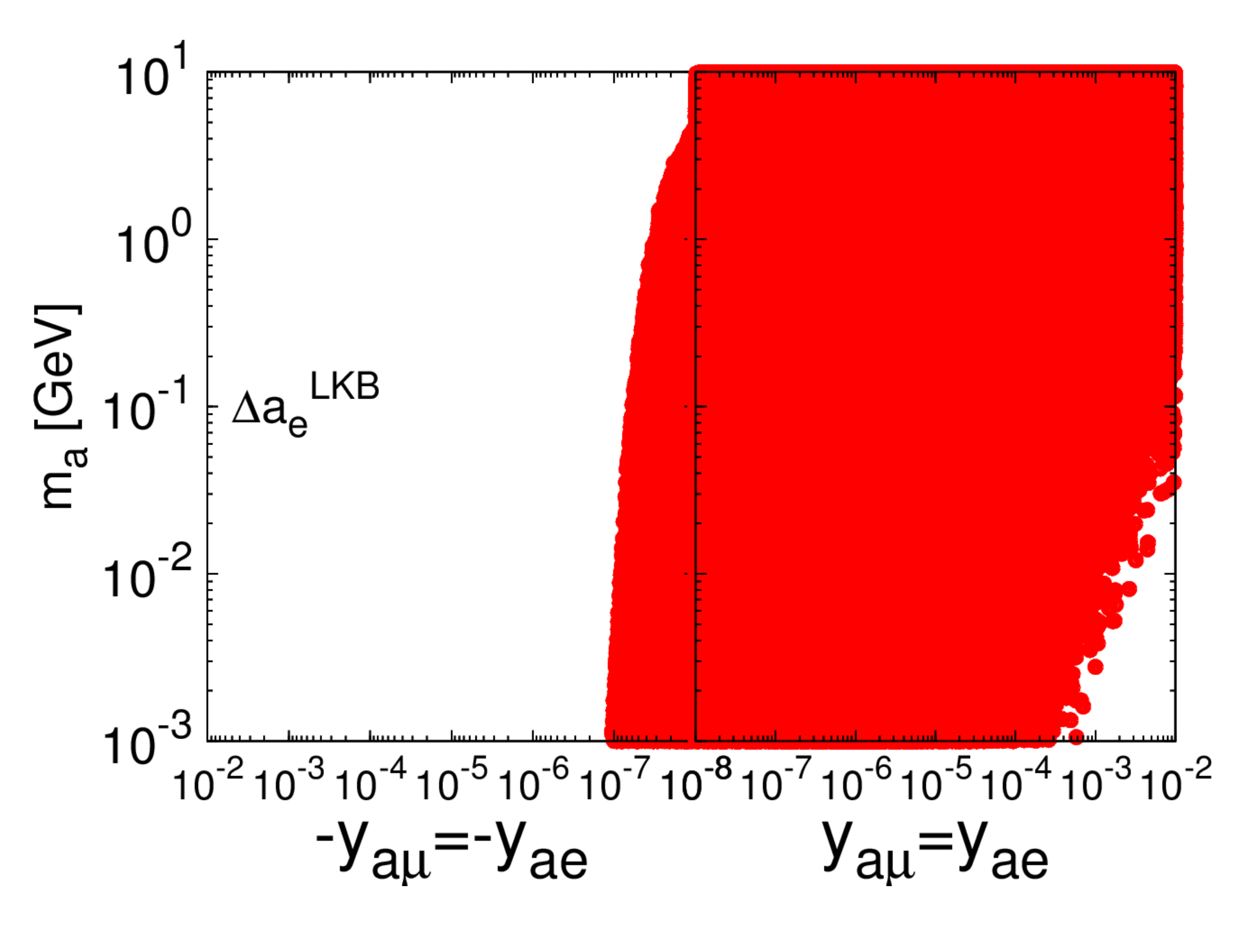}
\includegraphics[height=1.5in,angle=0]{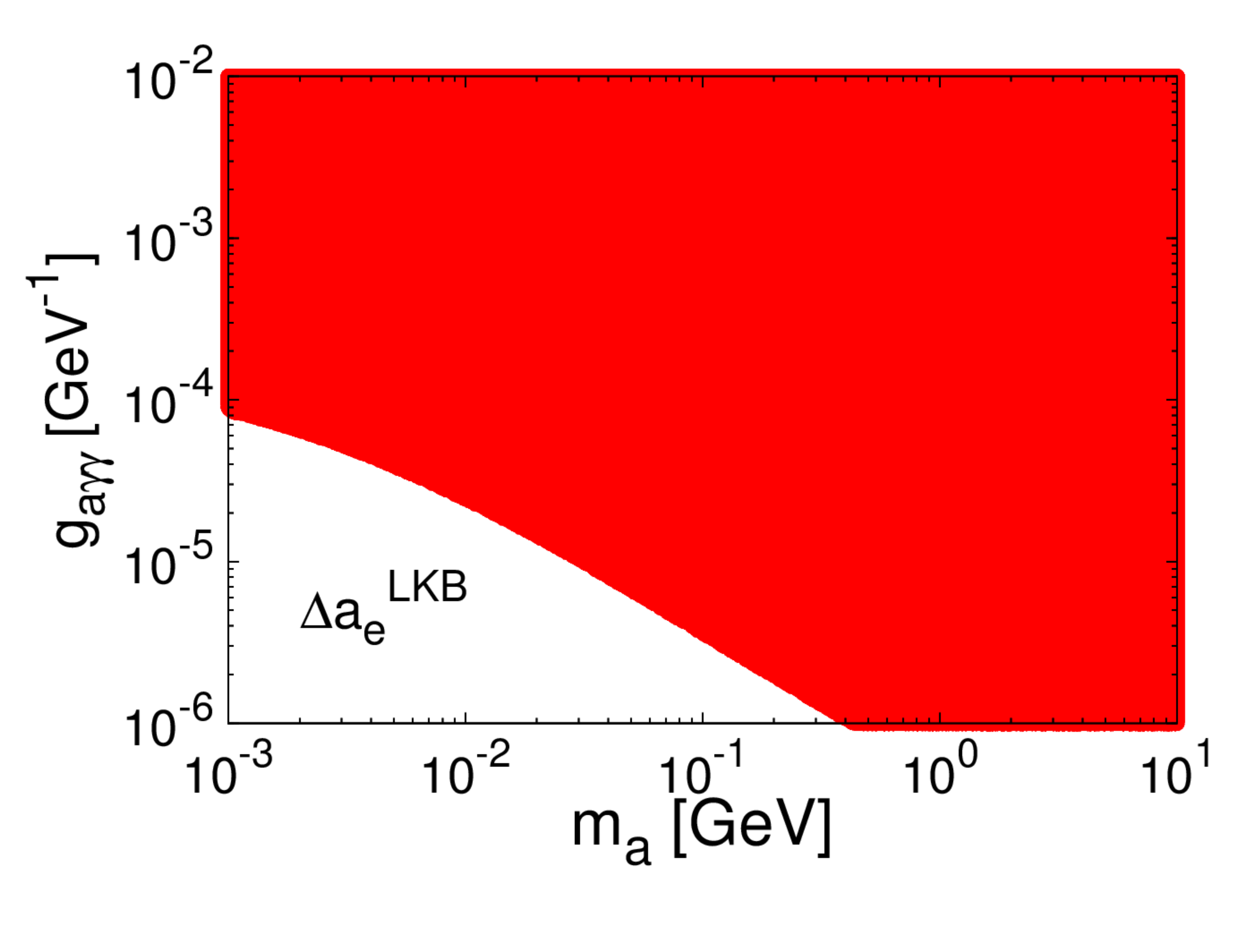}
\includegraphics[height=1.5in,angle=0]{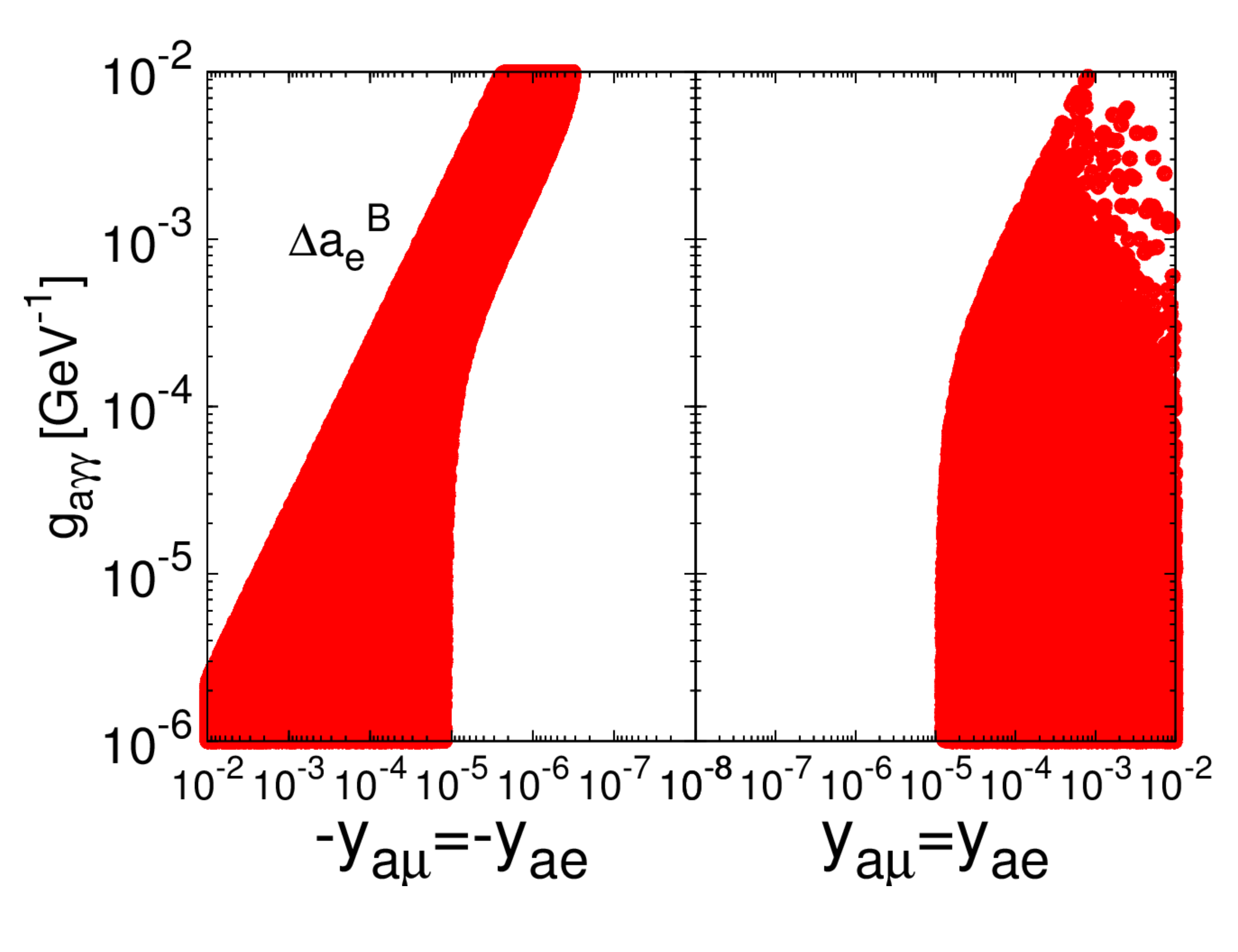}
\includegraphics[height=1.5in,angle=0]{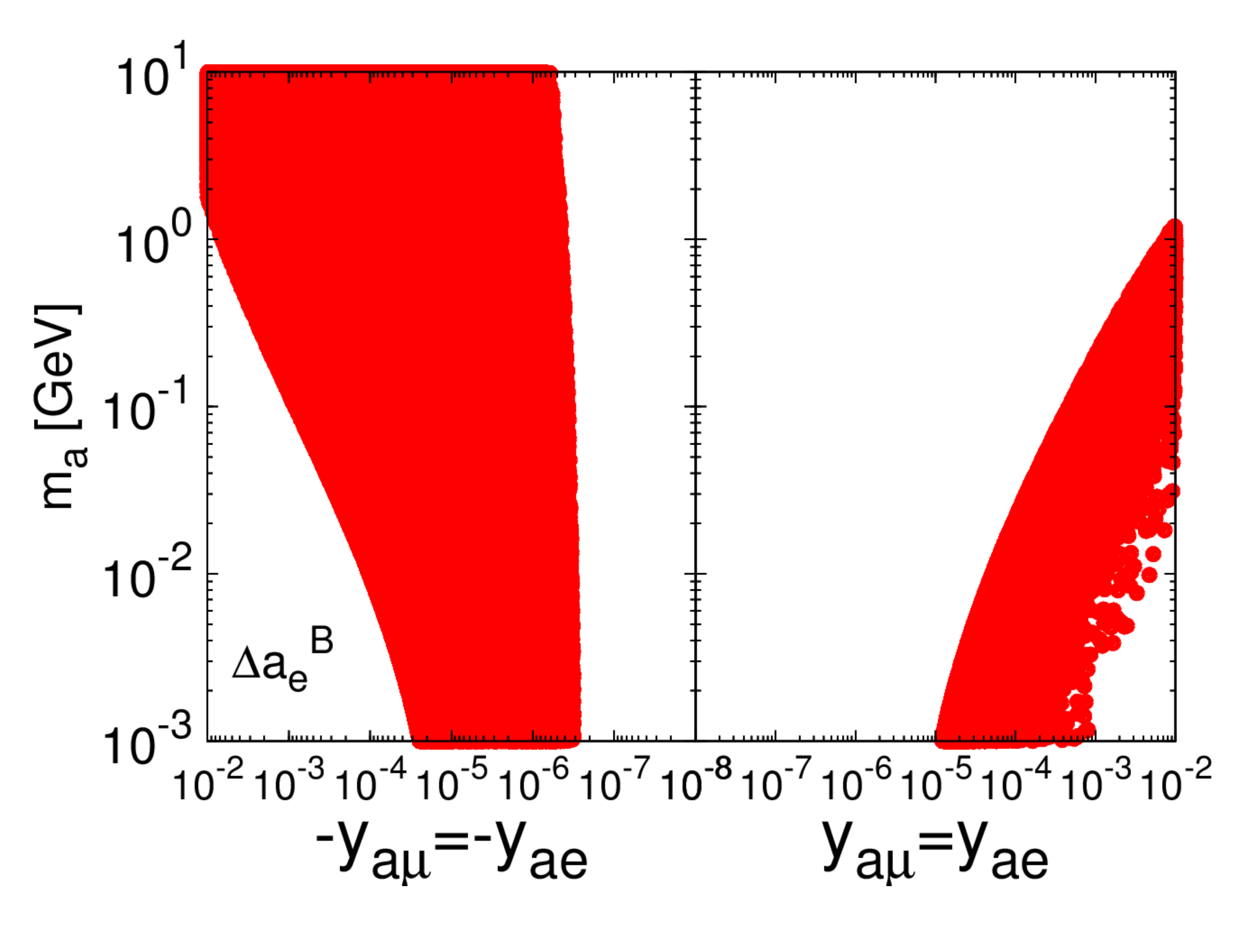}
\includegraphics[height=1.5in,angle=0]{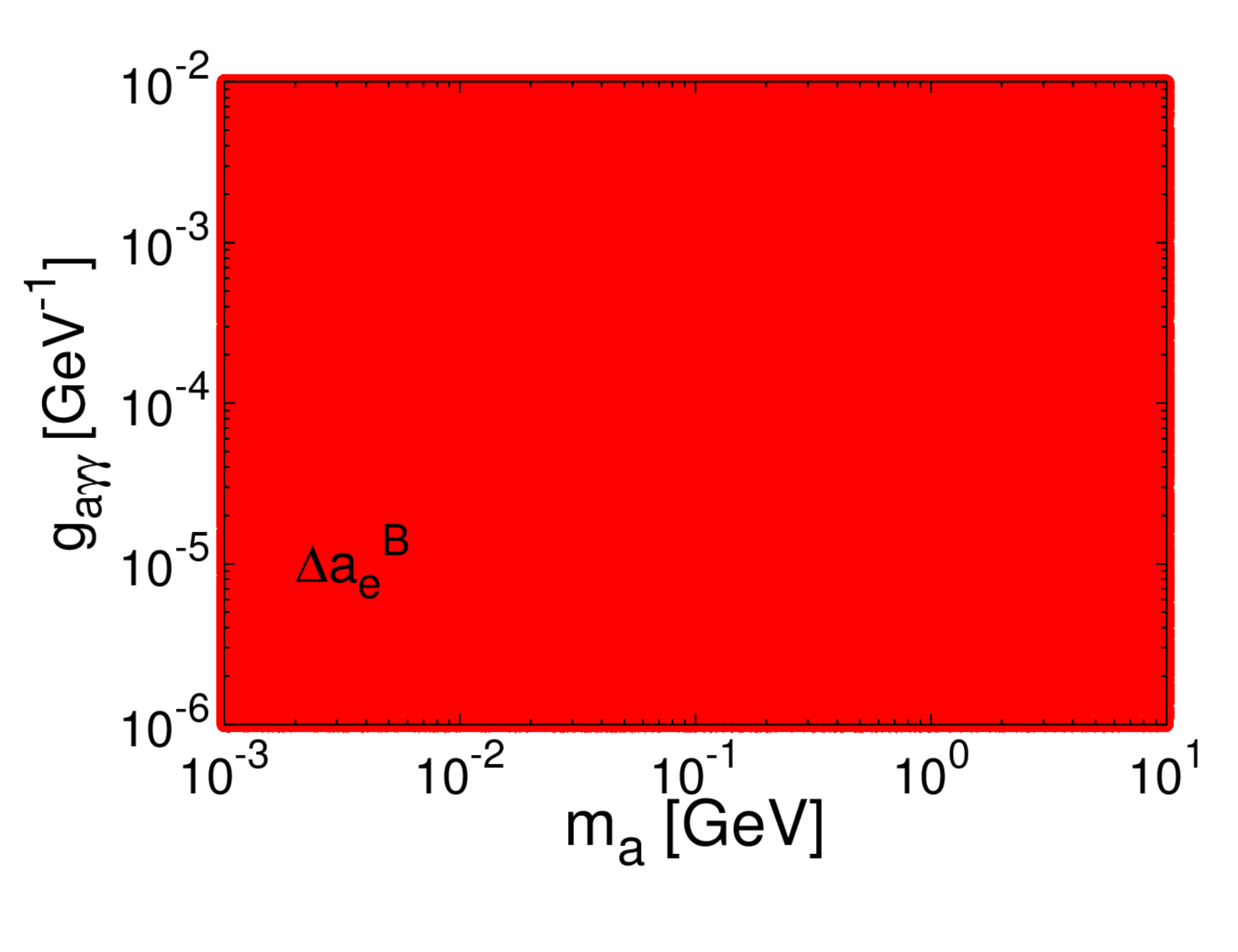}
\caption{\small \label{fig:scan2_g2_amu}
{\bf ALP-2:} 
 The 1$\sigma$ regions preferred by $\Delta a_\mu$ (upper panels), $\Delta a^{\rm LKB}_e$ (middle panels), and $\Delta a^{\rm B}_e$ (lower panels) for $y_{a\mu}=y_{ae}$.
}
\end{figure}

\begin{figure}[t]
\centering
\includegraphics[height=1.5in,angle=0]{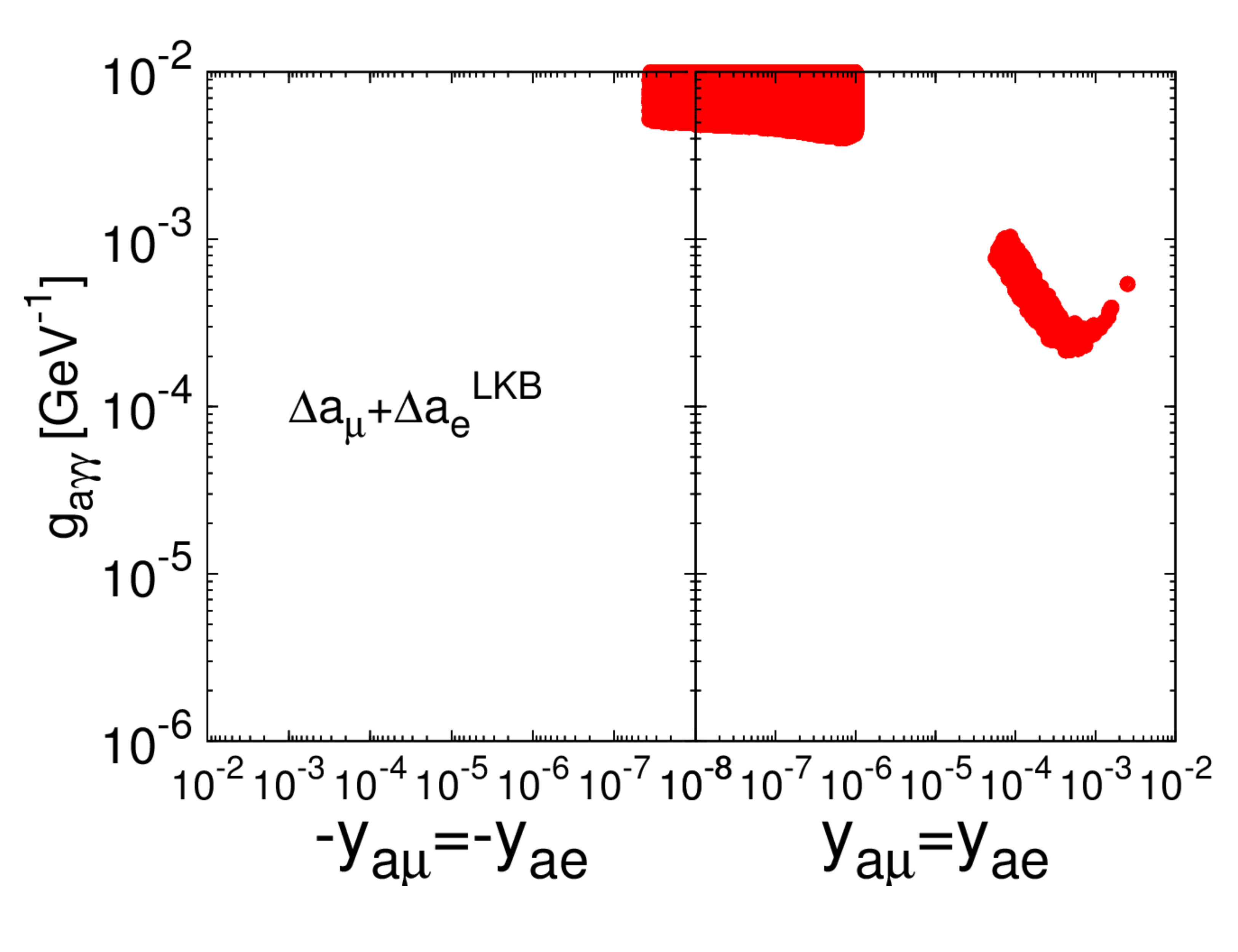}
\includegraphics[height=1.5in,angle=0]{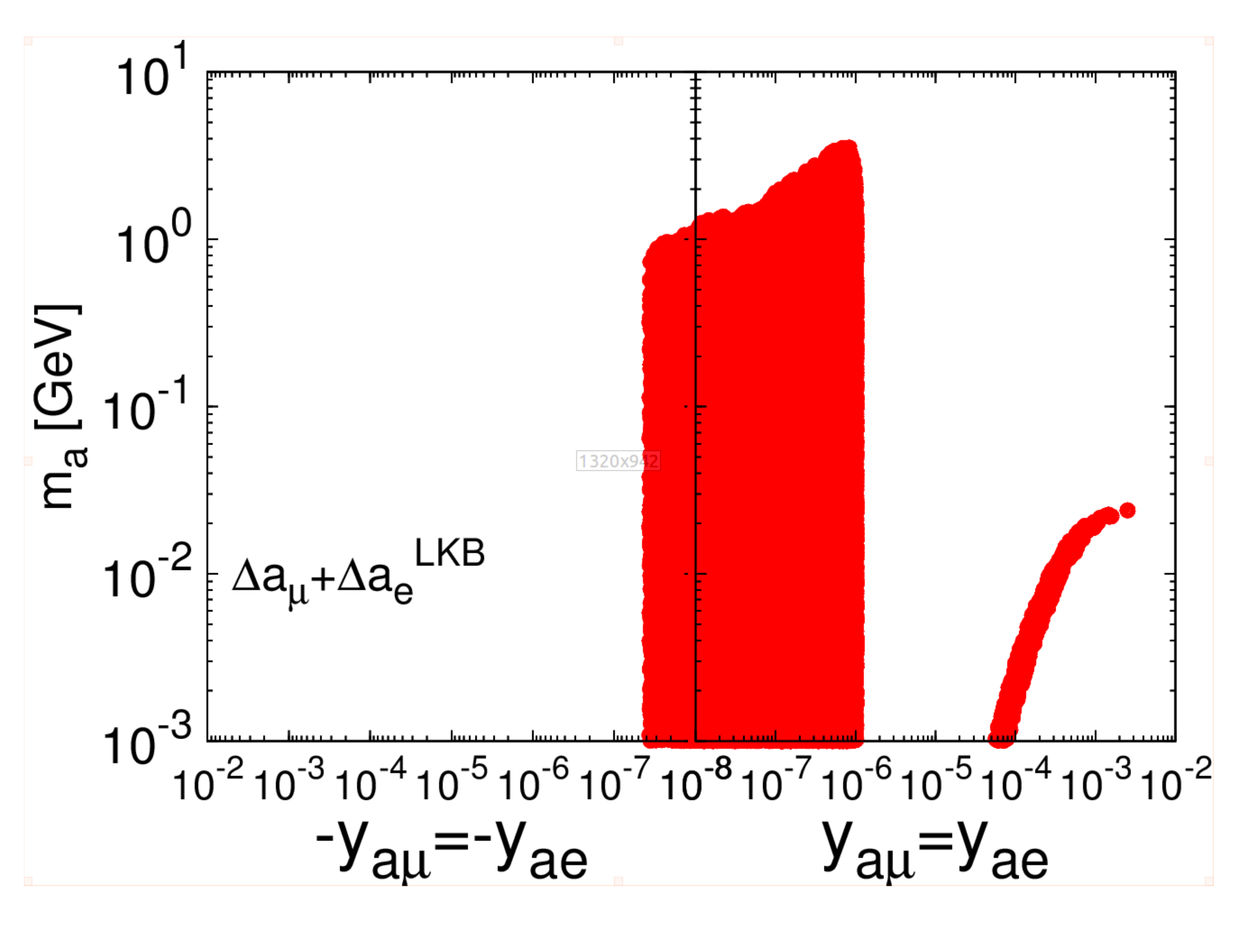}
\includegraphics[height=1.5in,angle=0]{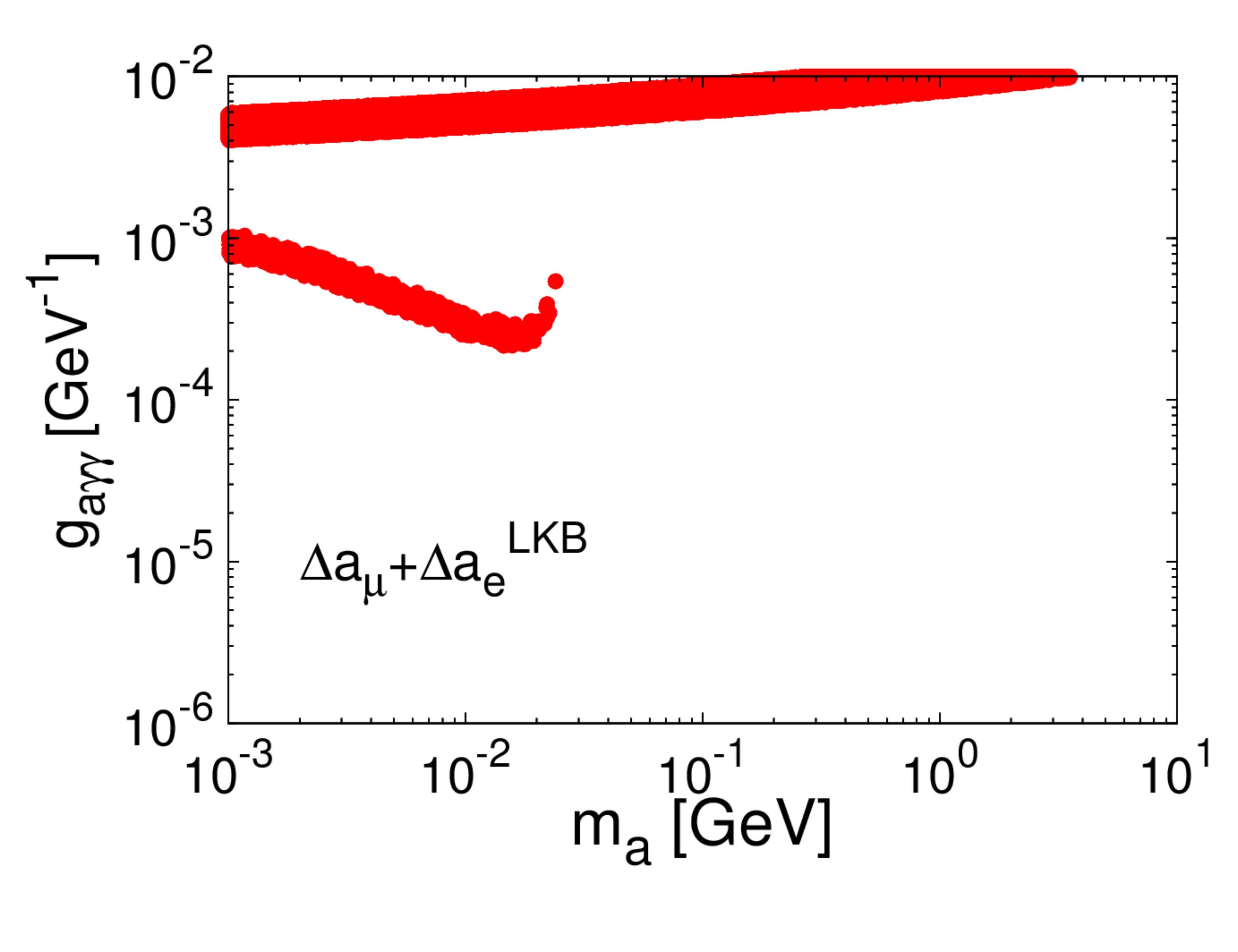}
\includegraphics[height=1.5in,angle=0]{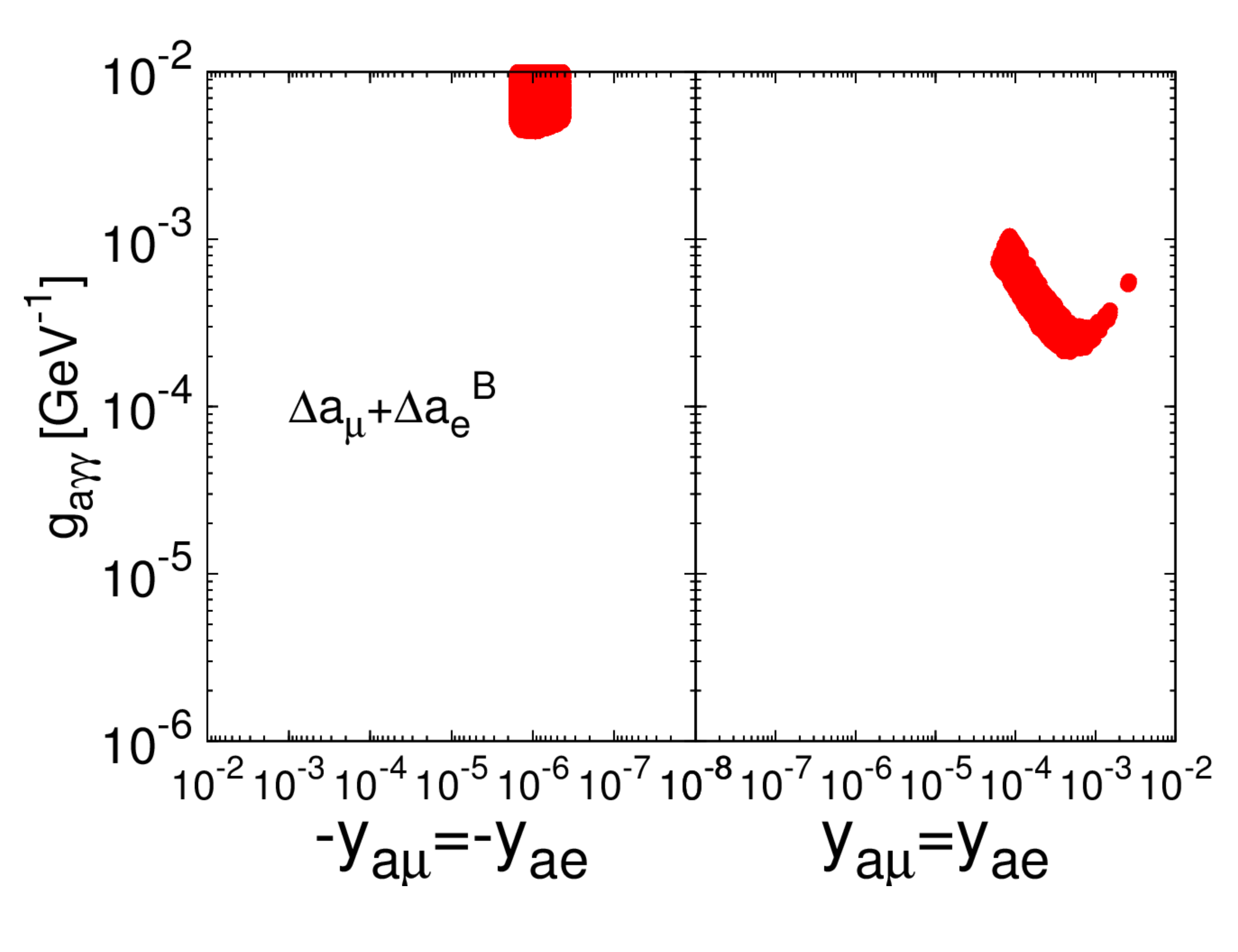}
\includegraphics[height=1.5in,angle=0]{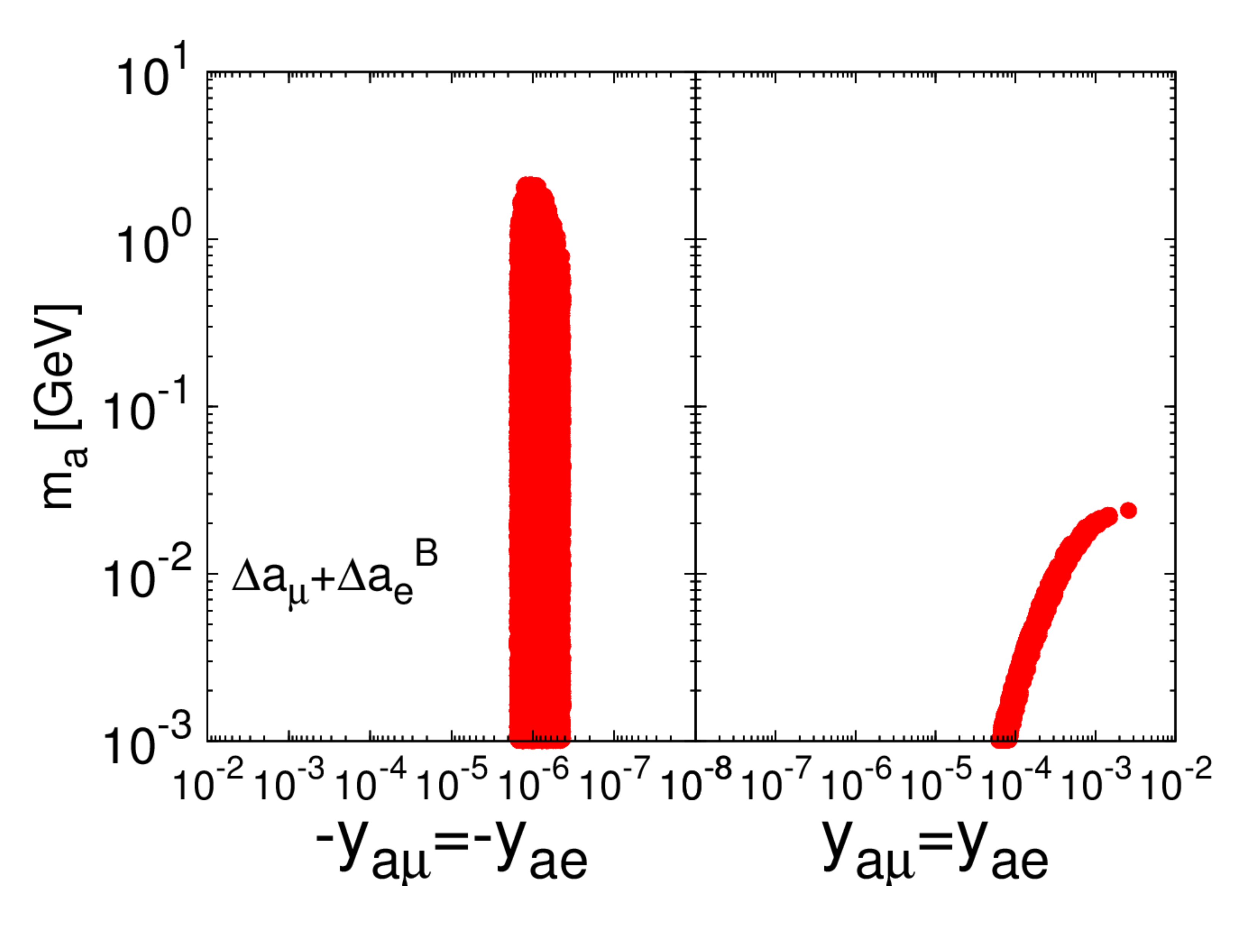}
\includegraphics[height=1.5in,angle=0]{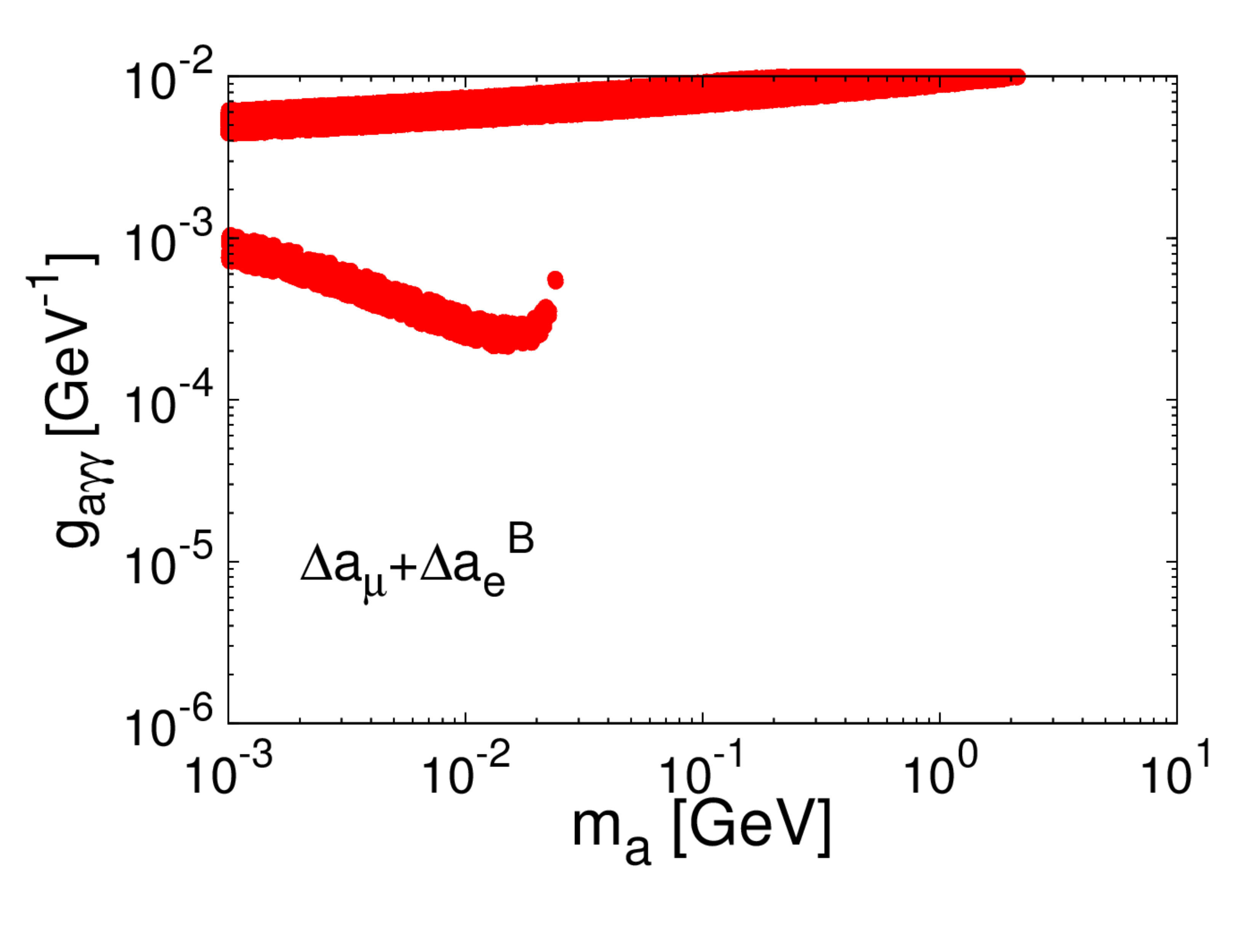}
\caption{\small \label{fig:scan2_g2_amu_ae}
{\bf ALP-2:} 
 The 1$\sigma$ allowed regions from a combined fit to $\Delta a_\mu$ and $\Delta a^{\rm LKB}_e$ (upper panels) and
 $\Delta a_\mu$ and $\Delta a^{\rm B}_e$ (lower panels), for $y_{a\mu}=y_{ae}$.
}
\end{figure}

Guided by the constraints mentioned above, we scan the parameter space in two scenarios which have the same number of free parameters:
\begin{itemize}
\item {\bf ALP-1:} We fix $m_a=0.2,1$~GeV, and vary $g_{a\gamma\gamma}$, $y_{a\mu}$, and $y_{ae}$. 
The results are shown in Figs.~\ref{fig:scan1_g2_amu} and \ref{fig:scan1_g2_amu_ae}. 
%
\item {\bf ALP-2:} We vary $m_a$, $g_{a\gamma\gamma}$ and $y_{a\mu}=y_{ae}$.
The results are shown in Figs.~\ref{fig:scan2_g2_amu} and \ref{fig:scan2_g2_amu_ae}.
\end{itemize}

 The minimum $\chi^2$ value in all cases is zero indicating that the deviations from the SM predictions can be exactly reproduced. 
 Since the $\chi^2$ distributions are very shallow around the minima, we do do not provide best-fit ALP points.

In the left panel of Fig.~\ref{fig:scan1_g2_amu}, the plateau for small Yukawa couplings arises from the LbL contribution. 
For large negative $y_{a\mu}$, 
the BZ and 1-loop contributions  interfere destructively with the LbL contribution which requires large $g_{a\gamma\gamma}$ values excluded by beam-dump experiments. For large positive $y_{a\mu}$ the BZ and LbL contributions interfere constructively so that the size of $g_{a\gamma\gamma}$ is reduced to fit the data. However, as $y_{a\mu}$ increases, the 1-loop contribution interferes destructively with the BZ and LbL contributions, which causes $g_{a\gamma\gamma}$ to rise again. A similar reasoning explains the structure of the $a_{e}^{\rm LKB}$
allowed region. A plateau does not appear in the allowed region for $\Delta a_{e}^{\rm B}$ because it is negative.




\section{Two-Higgs-doublet models}

\begin{figure}[t]
\centering
\includegraphics[height=1.5in,angle=0]{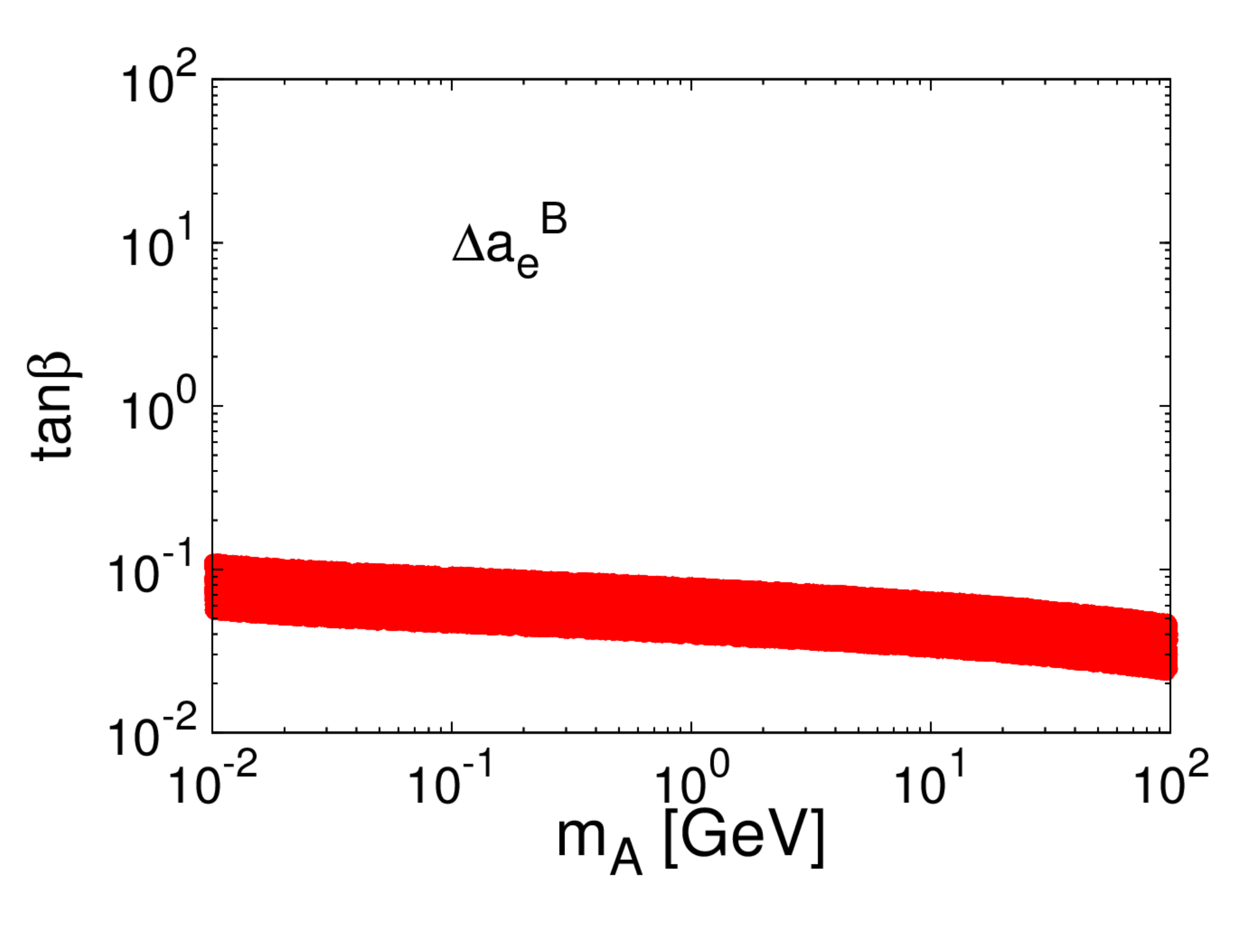}
\caption{\small \label{fig:I_2HDM}
{\bf Type-I 2HDM:} 
The 1$\sigma$ region preferred by  $\Delta a^{\rm B}_e$.
There is no solution for $\Delta a_\mu$ or $\Delta a^{\rm LKB}_e$ with $\chi^2<6.18$ (corresponding to $2\sigma$ for two parameters).
}
\end{figure}

\begin{figure}[t]
\centering
\includegraphics[height=1.5in,angle=0]{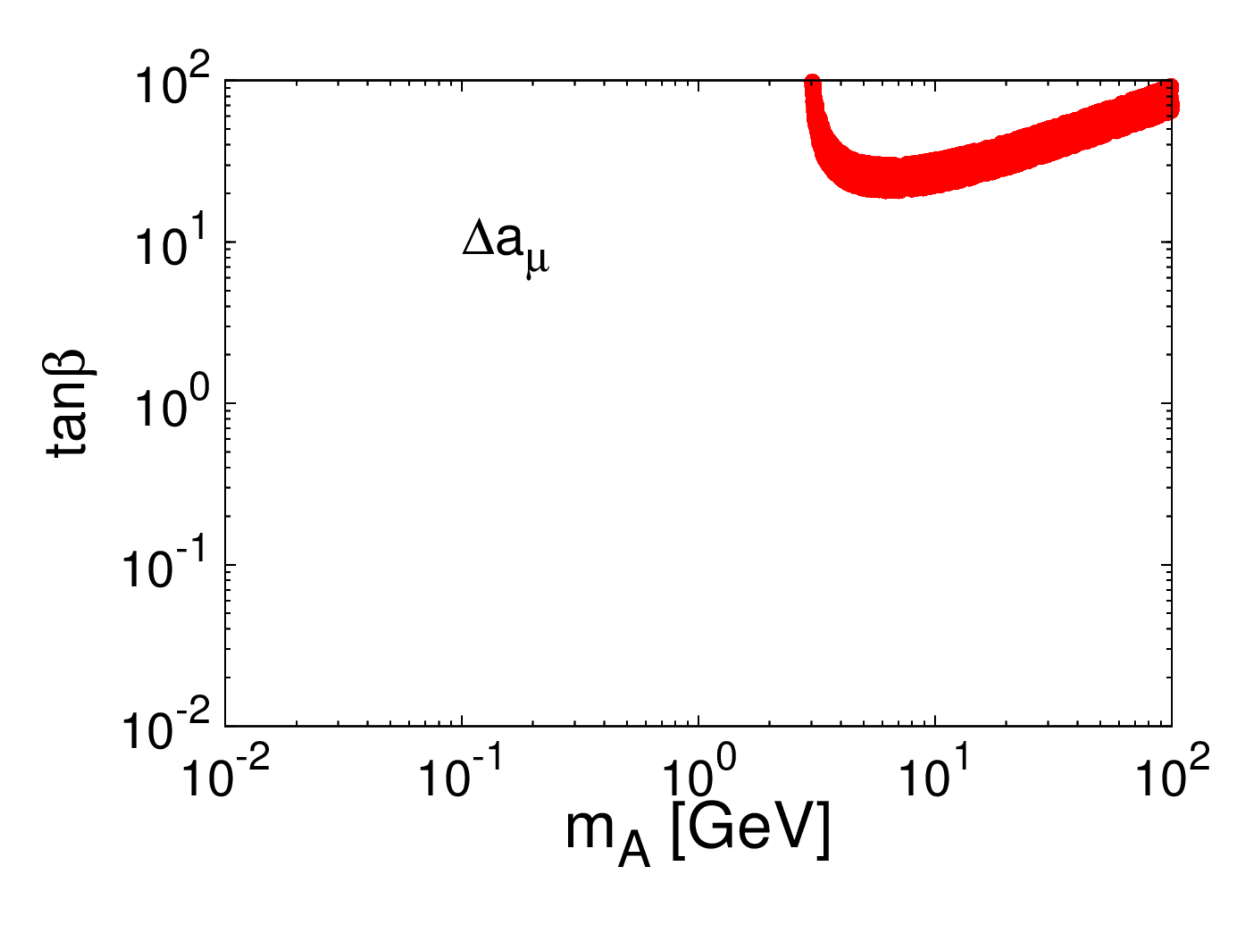}
\includegraphics[height=1.5in,angle=0]{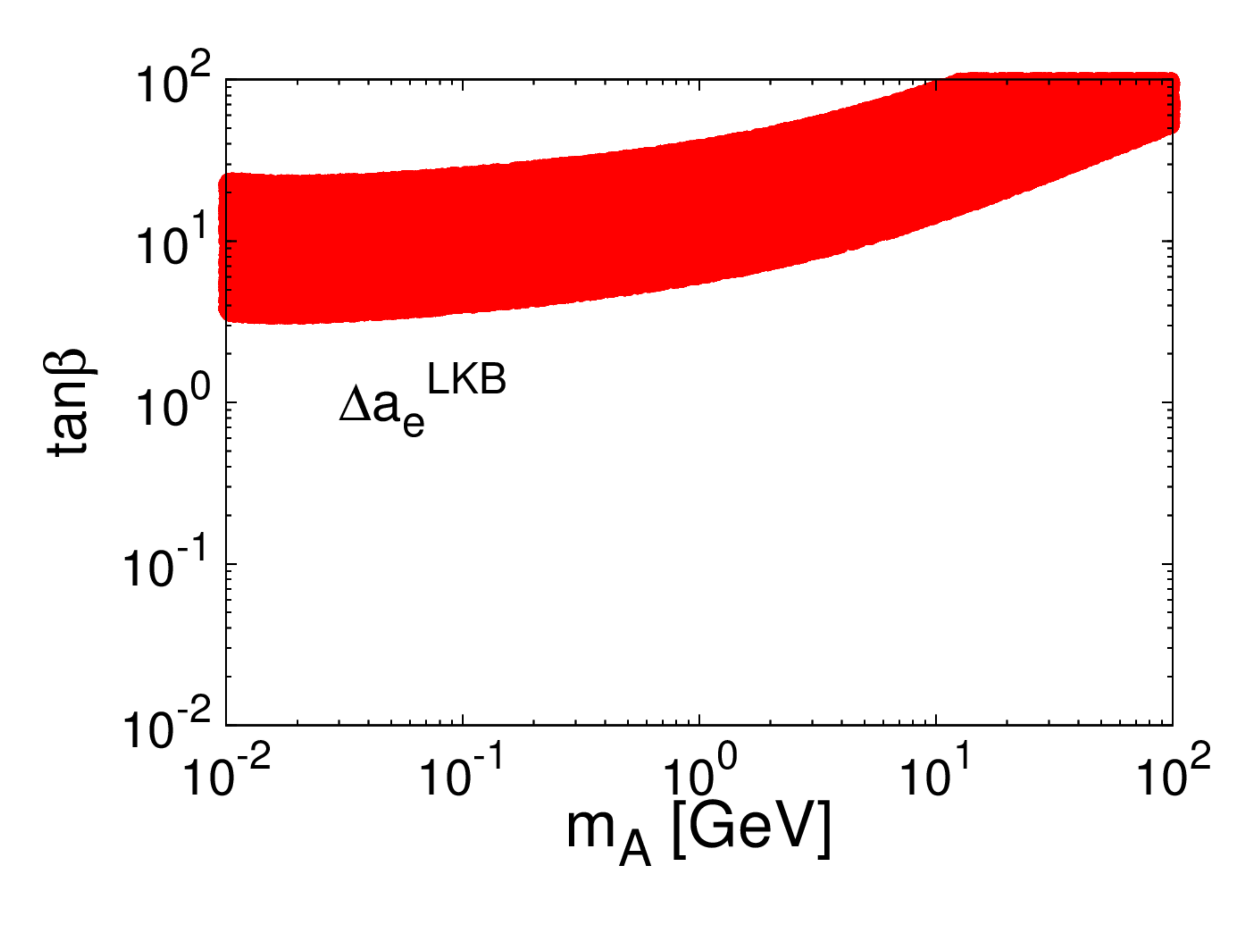}
\includegraphics[height=1.5in,angle=0]{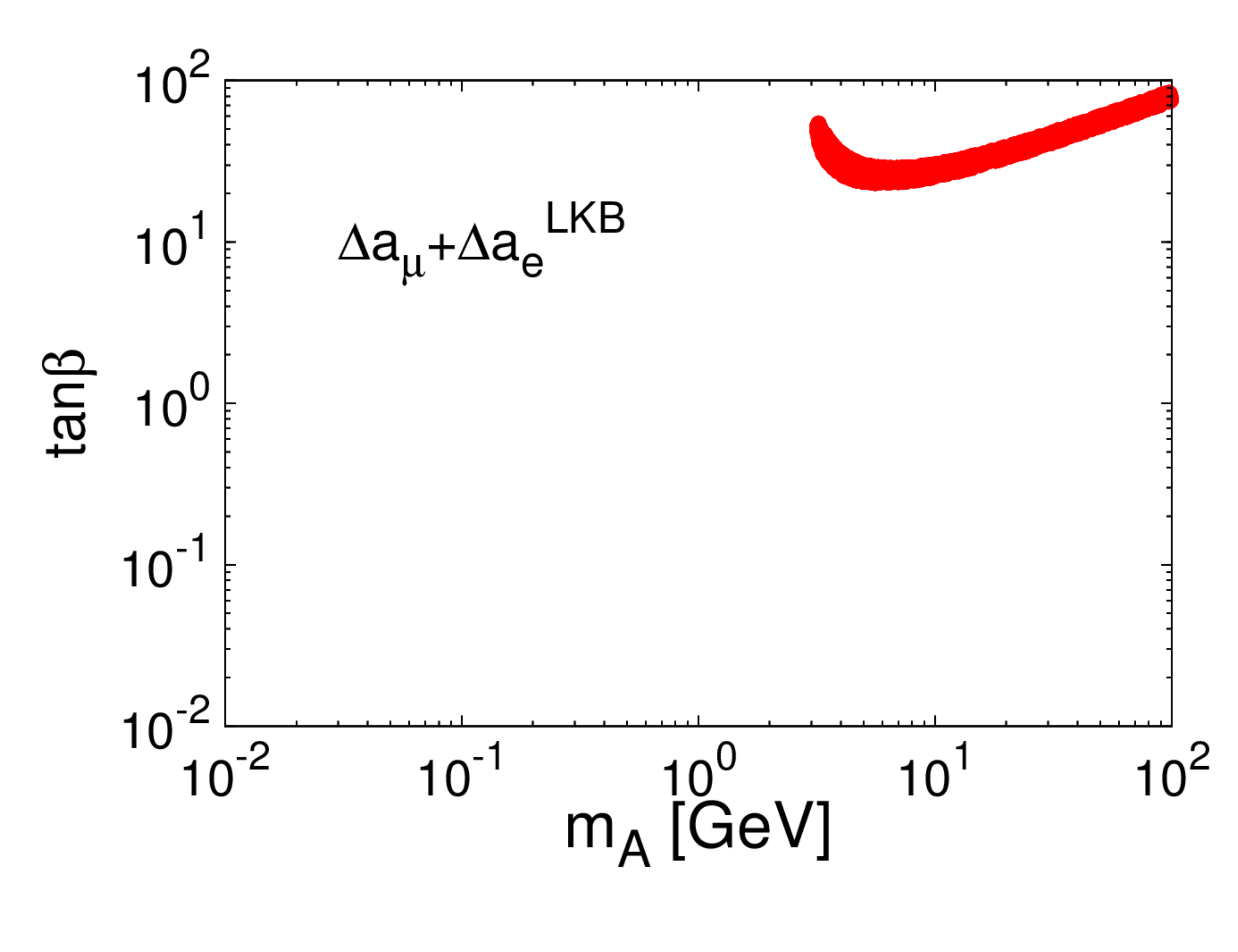}
\caption{\small \label{fig:II_2HDM}
{\bf Type-II 2HDM:} The 1$\sigma$ regions preferred by $\Delta a_\mu$, $\Delta a^{\rm {LKB}}_e$, and a combined fit of $\Delta a_\mu$ and $\Delta a^{\rm {LKB}}_e$.
There is no solution for $\Delta a^{\rm B}_e$ with $\chi^2<6.18$.
}
\end{figure}

\begin{figure}[t]
\centering
\includegraphics[height=1.5in,angle=0]{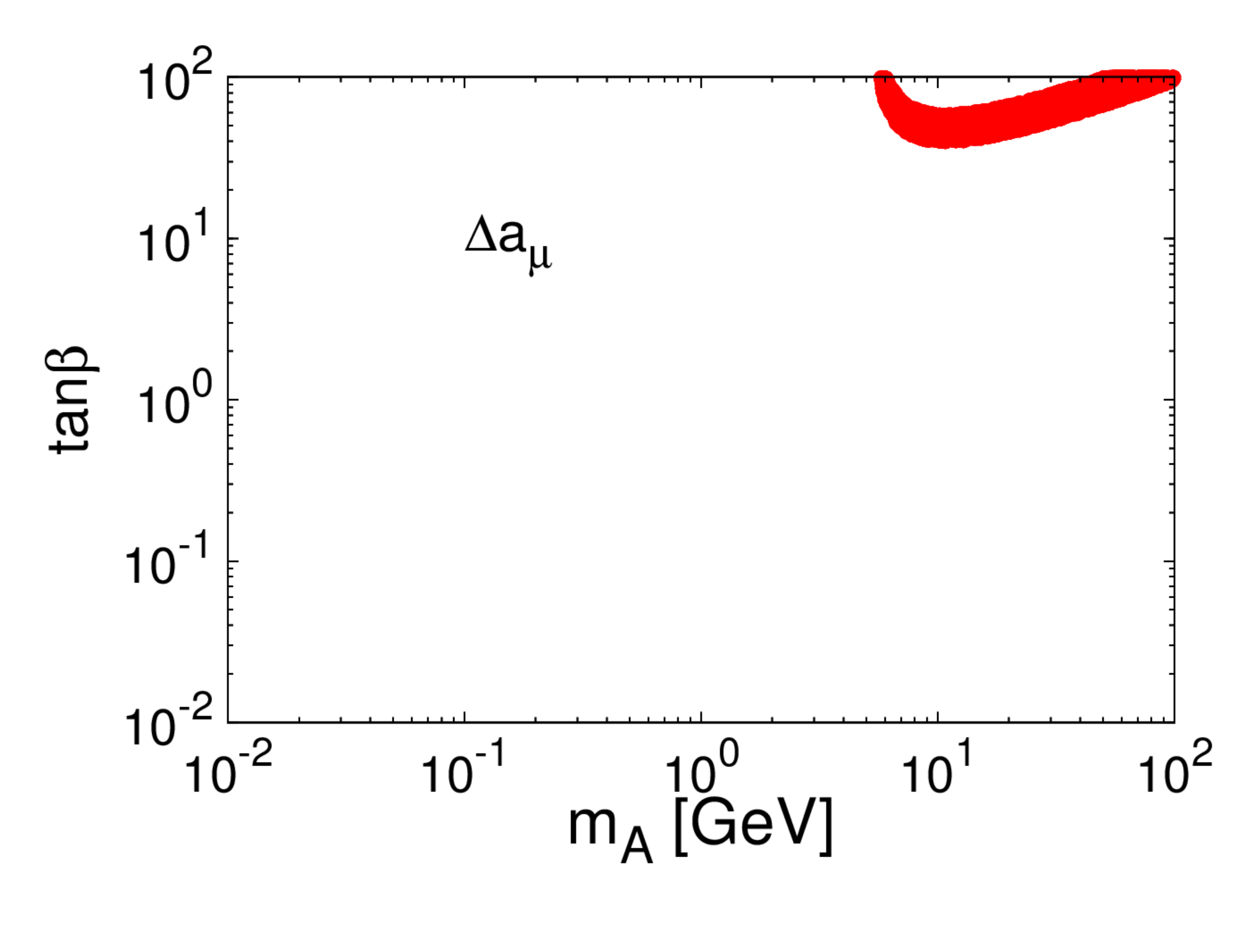}
\includegraphics[height=1.5in,angle=0]{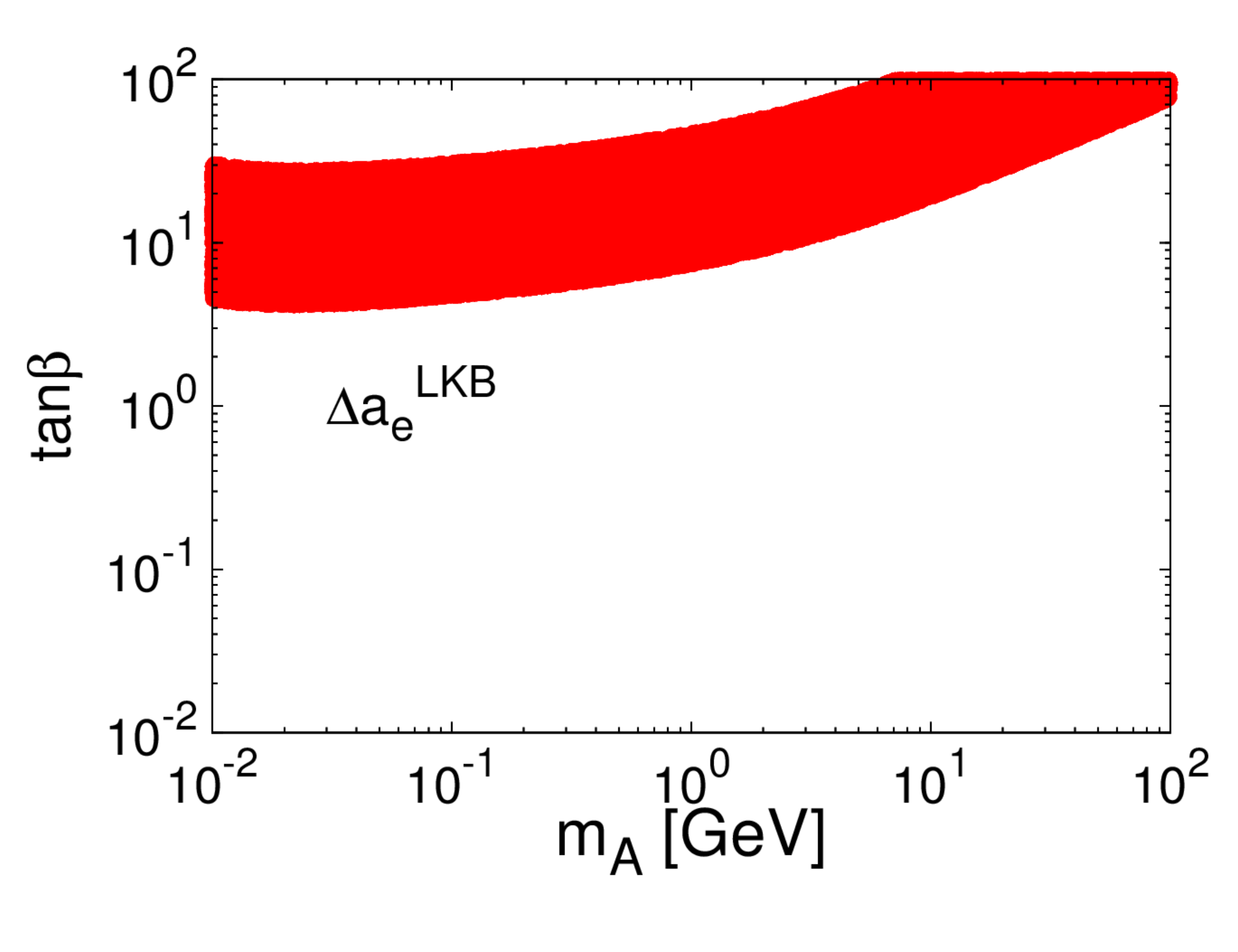}
\includegraphics[height=1.5in,angle=0]{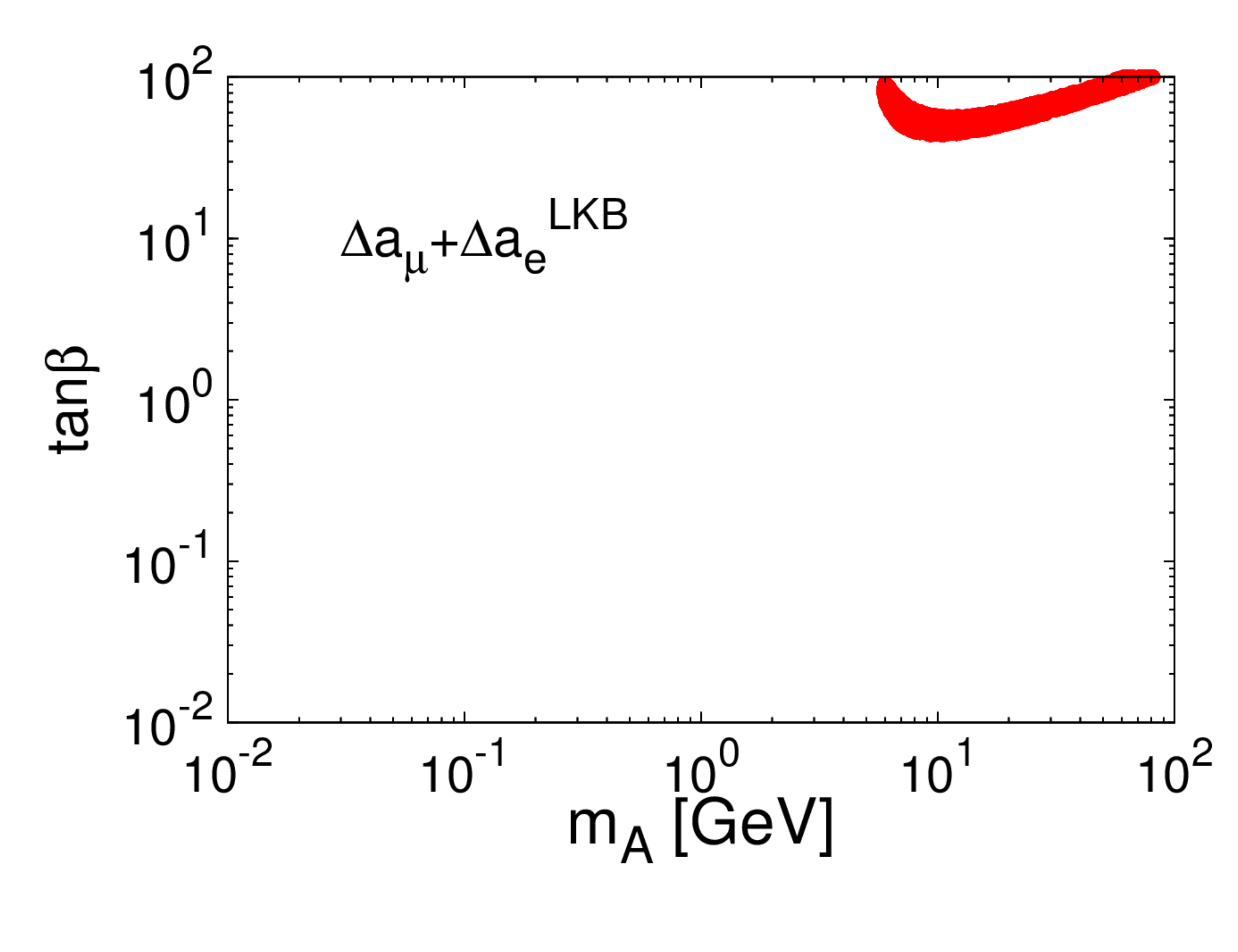}
\caption{\small \label{fig:X_2HDM}
{\bf Type-X 2HDM:} 
The 1$\sigma$ regions preferred by $\Delta a_\mu$, $\Delta a^{\rm {LKB}}_e$, and a combined fit of $\Delta a_\mu$ and $\Delta a^{\rm {LKB}}_e$.
There is no solution for $\Delta a^{\rm B}_e$ with $\chi^2<6.18$.
}
\end{figure}

\begin{figure}[t]
\centering
\includegraphics[height=1.5in,angle=0]{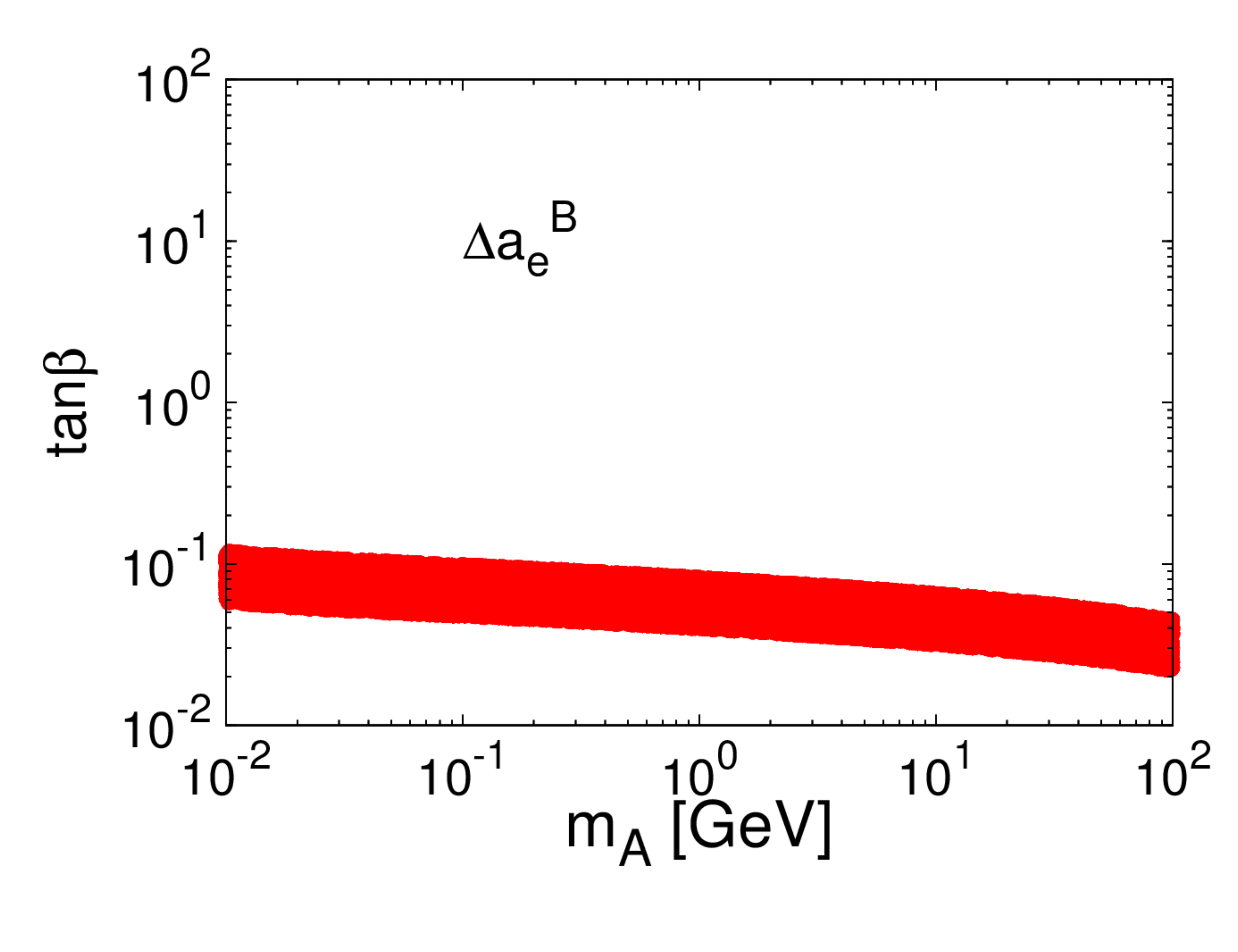}
\caption{\small \label{fig:Y_2HDM}
{\bf Type-Y 2HDM:} 
The 1$\sigma$ region preferred by  $\Delta a^{\rm B}_e$.
There is no solution for $\Delta a_\mu$ or $\Delta a^{\rm LKB}_e$ with $\chi^2<6.18$.
}
\end{figure}

We now consider two-Higgs-doublet models. 
In addition to the light Higgs $h$,
the scalar sector is comprised of a
heavy Higgs $H$, pseudoscalar $A$, 
and two charge Higgses $H^\pm$, which contribute to the electron and muon $g-2$
through either 1-loop triangle diagrams or two-loop BZ diagrams.
There are five relevant parameters, $m_A$, $m_H$, $m_{H^\pm}$, $\beta$, and $\alpha$.
The ratio of the vacuum expectation values of 
the two scalar doublets $\Phi_{1,2}$ defines $\tan \beta\equiv v_2/v_1$. The mixing 
between the CP-even neutral components $h_{1,2}$ of $\Phi_{1,2}$,
and the mass eigenstates $h,H$ is given by the angle $\alpha$~\cite{Branco:2011iw}:
\begin{eqnarray}
h &=& h_1 \sin \alpha - h_2 \cos \alpha\,, \nonumber \\
H &=& -h_1 \cos \alpha - h_2 \sin \alpha\,.
\end{eqnarray}
To satisfy the stringent constraints on flavor changing neutral currents, 
the up-type quarks, down-type quarks and leptons
must have Yukawa couplings to $\Phi_1$ or $\Phi_2$,  but not both.
This requirement leads to four types of 2HDMs:
Type-I (all fermions couple to $\Phi_2$),  
Type-II (only up-type quarks couple to $\Phi_2$),
Type-X (lepton-specific, in which only leptons couples to $\Phi_2$),
Type-Y (flipped, in which only down-type quarks couples to $\Phi_1$)~\cite{Branco:2011iw,Chun:2015xfx}.
The rare decay $b\to s \gamma$ requires $m_{H^\pm} \gtrsim 300$~GeV for $\tan \beta \gsim 2$ for Type-I and Type-X~\cite{Branco:2011iw} ($m_{H^\pm} \gtrsim 580$~GeV
 for Type-II and Type-Y~\cite{Misiak:2017bgg}),
which renders their contributions to $g-2$ subdominant.
Higgs precision measurements from ATLAS and CMS prefer $h$ to be SM-like with $\cos(\beta-\alpha) \to 0$, and $H$ decoupled.
After fixing $m_h=125.5$~GeV, we are left with only two parameters $m_A$ and $\tan\beta$ 
that affect $g-2$.
The Yukawa interactions in four types of 2HDMs are dictated by $\tan\beta$:
\begin{eqnarray}
-\mathcal{L}_{\rm 2HDM}^{\rm Yukawa} & = & \sum_{f=u,d,\ell}\frac{m_f}{v_{\rm EW}}
(y^h_f h \bar{f}f + y^H_f H\bar{f}f - iy^A_f \bar{f}\gamma_5 f) \nonumber \\
& +& \left[ \sqrt{2}V_{ud}H^+ \bar{u}
\left(\frac{m_u}{v_{\rm EW}}y^A_u P_L + \frac{m_d}{v_{\rm EW}}y^A_d P_R \right) d
+\frac{\sqrt{2}m_\ell}{v_{\rm EW}}H^+ \bar{\nu}P_R \ell + h.c. \right]\,,
\end{eqnarray}
with the normalized Yukawa couplings 
as listed in Table~\ref{tab:2HDM}.
The contribution to the anomalous magnetic moments is~\cite{Chun:2015xfx}
\begin{eqnarray}
&& a_{\ell, \rm 2HDM}= a_{\ell, \rm 2HDM}^{\rm 1-loop}+ a_{\ell, \rm 2HDM}^{\rm BZ}\,, \ \ \ {\text{where}}\\
&& a_{\ell, \rm 2HDM}^{\rm 1-loop}= \frac{G_F m^2_\ell}{4 \pi^2 \sqrt{2}}
\sum_{j=\left\lbrace h,H,A,H^\pm \right\rbrace}(y^j_\ell)^2 r^j_\ell f_j(r^j_\ell) \,, \nonumber \\
&&  a_{\ell, \rm 2HDM}^{\rm BZ}=
\frac{G_F m^2_\ell}{4\pi^2\sqrt{2}}\frac{\alpha}{\pi}
\sum_{i=\left\lbrace h,H,A \right\rbrace;f=\left\lbrace t,b,\tau \right\rbrace}
N^c_f Q^2_f y^i_\ell y^i_f r^i_f g_i(r^i_f)\,, \nonumber 
\end{eqnarray}
where $r^j_\ell\equiv m^2_\ell/m^2_j$, $r^i_f \equiv m^2_f/m^2_i$,
and $m_f$, $Q_f$, $N^c_f$ are the mass, electric charge and color factor for fermion $f$ in the loop.
The loop functions are
\begin{eqnarray}
f_{h,H}(r)&=&\int^1_0 dx \frac{x^2(2-x)}{1-x+rx^2} \,, \nonumber \\
f_{H^\pm}(r)&=&\int^1_0 dx \frac{-x(1-x)}{1-r(1-x)} \,, \nonumber \\
g_{h,H}(r)&=&\int^1_0 dx \frac{2x(1-x)-1}{x(1-x)-r}\ln \frac{x(1-x)}{r} \,, \nonumber \\
g_{A}(r)&=&\int^1_0 dx \frac{1}{x(1-x)-r}\ln \frac{x(1-x)}{r} \,,
\end{eqnarray}
and $f_A(r)$ is same as in Eq.~(\ref{eq:A_loop_function}).
The results of our statistical analysis 
by taking $\cos(\beta-\alpha)=0$ and $m_H=m_{H^\pm}=300$~GeV for Type-I and Type-X, and  $m_H=m_{H^\pm}=580$~GeV for Type-II and Type-Y, are shown in Figs.~\ref{fig:I_2HDM}, \ref{fig:II_2HDM}, \ref{fig:X_2HDM}, \ref{fig:Y_2HDM}, and Table~\ref{2HDM}.

Not surprisingly, the Type-I model does not reproduce the data because it is similar to the SM. Type-Y is similar to Type-I except that
the bottom quark contribution is enhanced. However, because of the lightness of the bottom quark, its contribution is not enough for the Type-Y model to explain the data.
For Type-X, the large value of $\tan \beta$ enhances the 
tau-lepton contribution to the two-loop BZ diagram. The contribution from the bottom-quark in the BZ diagram provides a further enhancement in the Type-II model. Note that the Type-II parameters needed are excluded by constraints from $B_s \to \mu^+\mu^-$
and searches for $Z \to b\bar{b}A(b\bar{b})$~\cite{Chun:2015xfx}.

\begin{table}[t]
\caption{\small \label{tab:2HDM}
The normalized Yukawa couplings for the two-Higgs-doublet models~\cite{Branco:2011iw} .
}
\begin{adjustbox}{width=\textwidth}
\begin{tabular}{c|ccccccccc}
\hline
\hline
    & $y^A_u$ & $y^A_d$ & $y^A_\ell$ 
    & $y^H_u$ & $y^H_d$ & $y^H_\ell$ 
    & $y^h_u$ & $y^h_d$ & $y^h_\ell$  \\
\hline
Type-I     
    & $\cot \beta$ & $-\cot \beta$ & $-\cot \beta$  
    & $\frac{\sin \alpha}{\sin \beta}$ & $\frac{\sin \alpha}{\sin \beta}$ & $\frac{\sin \alpha}{\sin \beta}$
    & $\frac{\cos \alpha}{\sin \beta}$ & $\frac{\cos \alpha}{\sin \beta}$ & $\frac{\cos \alpha}{\sin \beta}$    \\
Type-II    
    & $\cot \beta$ & $\tan \beta$ & $\tan \beta$  
    & $\frac{\sin \alpha}{\sin \beta}$ & $\frac{\cos \alpha}{\cos \beta}$ & $\frac{\cos \alpha}{\cos \beta}$
    & $\frac{\cos \alpha}{\sin \beta}$ & $-\frac{\sin \alpha}{\cos \beta}$ & $-\frac{\sin \alpha}{\cos \beta}$    \\
Type-X    
    & $\cot \beta$ & $-\cot \beta$ & $\tan \beta$  
    & $\frac{\sin \alpha}{\sin \beta}$ & $\frac{\sin \alpha}{\sin \beta}$ & $\frac{\cos \alpha}{\cos \beta}$
    & $\frac{\cos \alpha}{\sin \beta}$ & $\frac{\cos \alpha}{\sin \beta}$ & $-\frac{\sin \alpha}{\cos \beta}$    \\
Type-Y   
    & $\cot \beta$ & $\tan \beta$ & $-\cot \beta$  
    & $\frac{\sin \alpha}{\sin \beta}$ & $\frac{\cos \alpha}{\cos \beta}$ & $\frac{\sin \alpha}{\sin \beta}$
    & $\frac{\cos \alpha}{\sin \beta}$ & $-\frac{\sin \alpha}{\cos \beta}$ & $\frac{\cos \alpha}{\sin \beta}$    \\
\hline
\hline
\end{tabular}
\end{adjustbox}
\end{table}

\begin{table}[t]
\caption{\small \label{tab:min_2HDM}
Best-fit points, their contributions to the anomalous magnetic moments, and minimum $\chi^2$ values for the 2HDMs.
}
\begin{adjustbox}{width=\textwidth}
\begin{tabular}{c|cc|cc|cc|cc}
\hline
\hline
    & \multicolumn{2}{c}{Type-I}  & \multicolumn{2}{c}{Type-II} & \multicolumn{2}{c}{Type-X} & \multicolumn{2}{c}{Type-Y}  \\
\hline
& LKB & B & LKB & B & LKB & B & LKB & B  \\
\hline
$m_A/{\rm GeV}$     
            & 98.5 & 98.5 & 3.44 & 99.2 & 6.52 & 71.5 & 99.8  & 99.8  \\
$\tan\beta$     
            & 99.4 & 99.4 & 40.9 & 88.1 & 68.9 & 99.3 & 94.3 & 94.3 \\
\hline
$ a_\mu$
            & $-1.5\times 10^{-11}$ & $-1.5\times 10^{-11}$ 
            & $2.5\times 10^{-9}$ & $2.4\times 10^{-9}$ 
            & $2.5\times 10^{-9}$ & $2.5\times 10^{-9}$ 
            & $-1.6\times 10^{-11}$ & $-1.6\times 10^{-11}$ \\
$a_e$
            & $-3.6\times 10^{-16}$ & $-3.6\times 10^{-16}$ 
            & $4.8\times 10^{-13}$  & $5.8\times 10^{-14}$ 
            & $4.9\times 10^{-13}$  & $6.5\times 10^{-14}$ 
            & $-3.7\times 10^{-16}$ & $-3.7\times 10^{-16}$ \\
\hline
$\chi^2_{\rm LKB}$ & 20.9 & -    & $0$ & - & $0$ & - & 20.9 & - \\
$\chi^2_{\rm B}$   & -    & 24.3 & - & 6.81 & - & 6.90 & - & 24.3 \\
\hline
\hline
\end{tabular}
\end{adjustbox}
\label{2HDM}
\end{table}

\section{Leptoquarks}

We consider a scalar leptoquark $S_1\sim ({\bf 3},{\bf 1},-1/3)$ and a doublet leptoquark 
$(R_2)^T=(R^{5/3}_2,R^{2/3}_2)\sim ({\bf 3},{\bf 2},7/6)$.
Their couplings to quarks and leptons are specified in the ``up-type" mass-diagonal basis 
because the ``down-type" basis would violate constraints from $\mu \to e \gamma$~\cite{Dorsner:2020aaz}.
Then, the CKM matrix appears in the couplings with down-type quarks, 
and the interaction Lagrangian is~\cite{Bigaran:2020jil}
\begin{eqnarray}
&& \mathcal{L}^{S_1}\supset y^{SLQ}_{ij}
\left[ \bar{e}^c_{L,i}u_{L,j}-V^{\rm CKM}_{jk} \bar{\nu}^c_{L,i}d_{L,k} \right]S^\dagger_1
+y^{Seu}_{ij}\bar{e}^c_{R,i}u_{R,j}S^\dagger_1+h.c.\,, \nonumber \\
&& \mathcal{L}^{R_2}\supset
y^{RLu}_{ij}\left[\bar{\nu}_{L,i}u_{R,j}R^{2/3,\dagger}_2-\bar{e}_{L,i}u_{R,j} R^{5/3,\dagger}_2 \right]
+y^{ReQ}\bar{e}_{R,i}\left[u_{L,j}R^{5/3,\dagger}_2+V^{\rm CKM}_{jk}d_{L,k}R^{2/3,\dagger}_2 \right]+h.c. \nonumber \\
\end{eqnarray}
$S_1$ and $R^{5/3}_2$ have left- and right-handed 
couplings to the charge leptons and up-type quarks, and give a large contribution to $a_\ell$
under the condition $m_q \gg m_\ell$.
We neglect the contribution of $R^{2/3}_2$, which only has right-handed couplings 
to down quarks.
In the limit, $m_q \ll m_{\rm LQ}$, 
\begin{eqnarray}
\label{eq:LQ_g2}
a_{\ell, S_1} \simeq - \frac{m_\ell m_q}{4\pi^2 m^2_{S_1}}
\left[ \frac{7}{4}-2 \log\left( \frac{m_{S_1}}{m_q} \right) \right]
{\rm Re}(y^{L*}_{\ell q} y^R_{\ell q})\,, \nonumber \\
a_{\ell, R_2} \simeq \frac{m_\ell m_q}{4\pi^2 m^2_{R_2}}
\left[ \frac{1}{4}-2 \log\left( \frac{m_{R_2}}{m_q} \right) \right]
{\rm Re}(y^{L*}_{\ell q} y^R_{\ell q})\,,
\end{eqnarray}
where for $S_1$,
$y^R_{ij}\equiv y^{Seu}_{ij}$ and $y^L_{ij}\equiv y^{SLQ}_{ij}$; 
and for $R^{5/3}_2$,
$y^R_{ij}\equiv -y^{RLu}_{ij}$ and $y^L_{ij}\equiv y^{ReQ}_{ij}$.
Note that $a_\ell$ requires both non-vanishing 
left- and right-handed Yukawa couplings.
There is freedom to choose the texture of the Yukawa couplings $y^{L,R}_{ij}$.
According to Eq.~(\ref{eq:LQ_g2}), 
heavier fermions contribute more to $a_\ell$,
so we ignore the $u$ quark and $\tau$ lepton.
The remaining couplings are $y^{L,R}_{ec,\,et,\,\mu c,\, \mu t}$. 
However, if both $y_{e t}$ and $y_{\mu t}$ are non-zero,
the $m_t$ enhancement of $\mu \to e \gamma$ becomes incompatible 
with observation~\cite{Bigaran:2020jil}.
A non-zero $y_{\mu c}$ allows the LQ to couple to neutrinos and the $s$ quark,
thereby inducing $K^+ \to \pi^+ \nu \nu$ through CKM mixing.
To obey these constraints we set $y_{et}=y_{\mu c}=0$. 
Finally, we have the Yukawa couplings,
\begin{eqnarray}
y^L_{ij} \sim \left(
\begin{array}{ccc}
0 & y^L_{ec} & 0 \\[2mm]
0 & 0        & y^L_{\mu t} \\[2mm]
0 & 0        & 0
\end{array}
\right),~~~~
y^R_{ij}\sim \left(
\begin{array}{ccc}
0 & y^R_{ec} & 0 \\[2mm]
0 & 0        & y^R_{\mu t} \\[2mm]
0 & 0        & 0
\end{array}
\right)\,.
\end{eqnarray}

\begin{figure}[t]
\centering
\includegraphics[height=1.5in,angle=0]{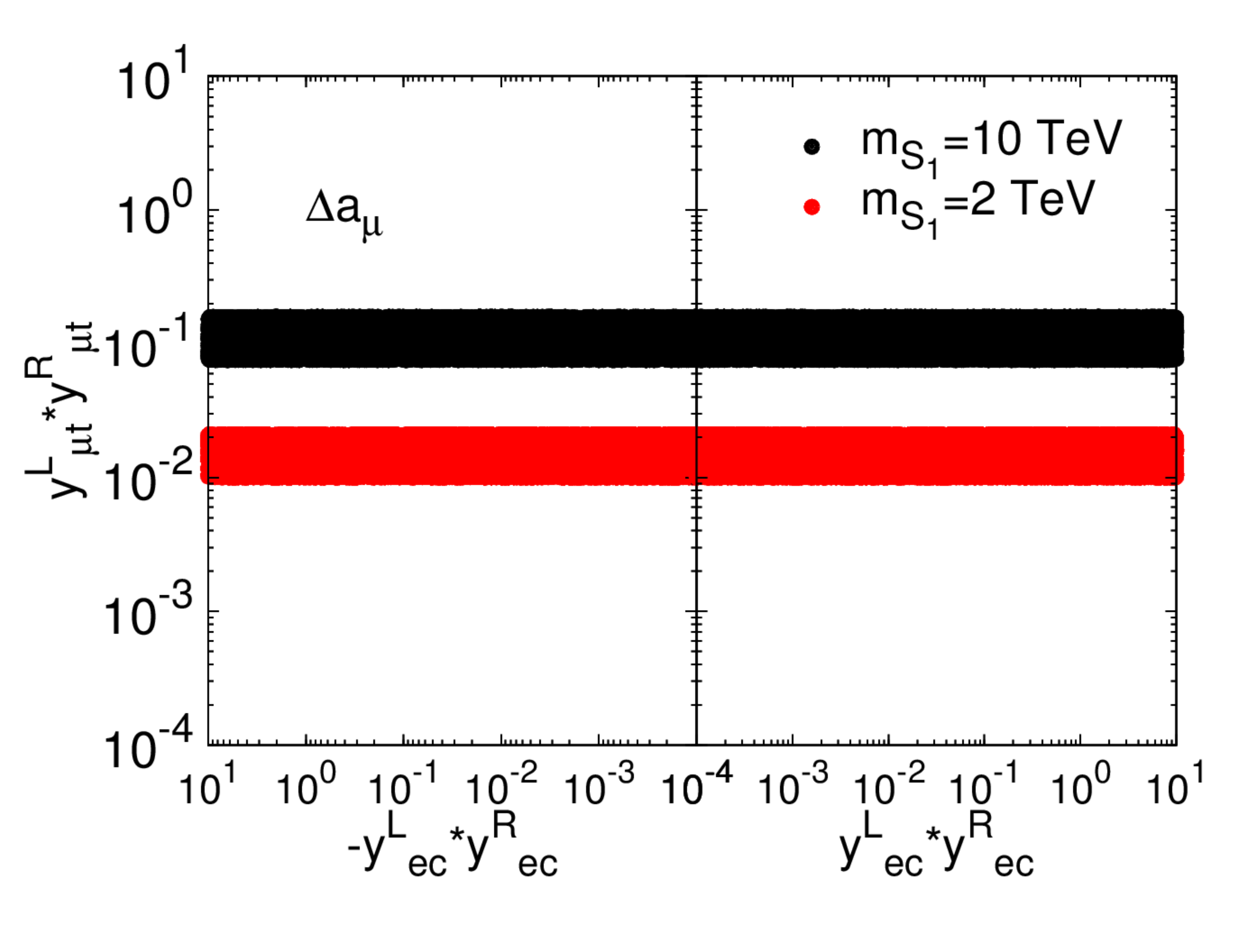}
\includegraphics[height=1.5in,angle=0]{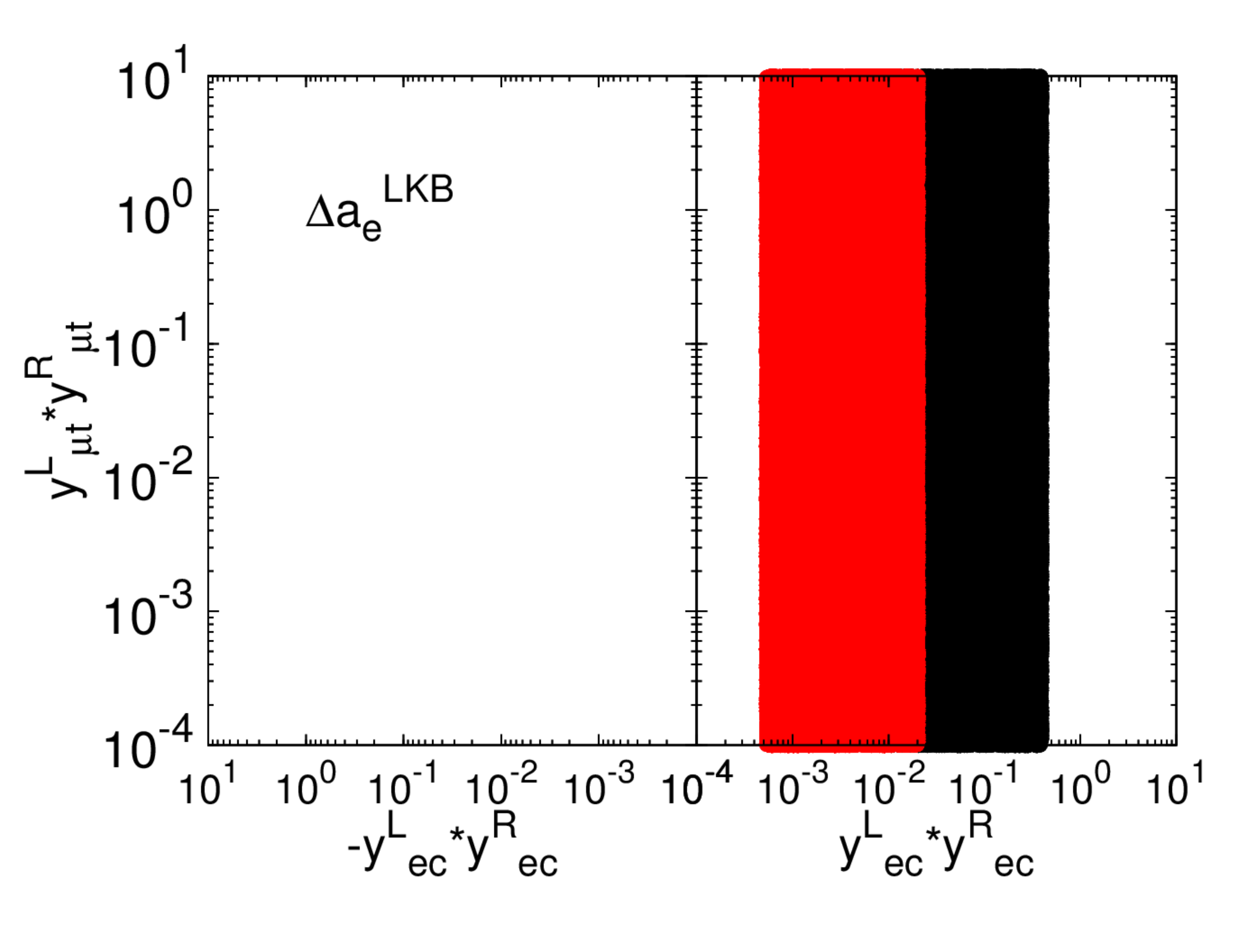}
\includegraphics[height=1.5in,angle=0]{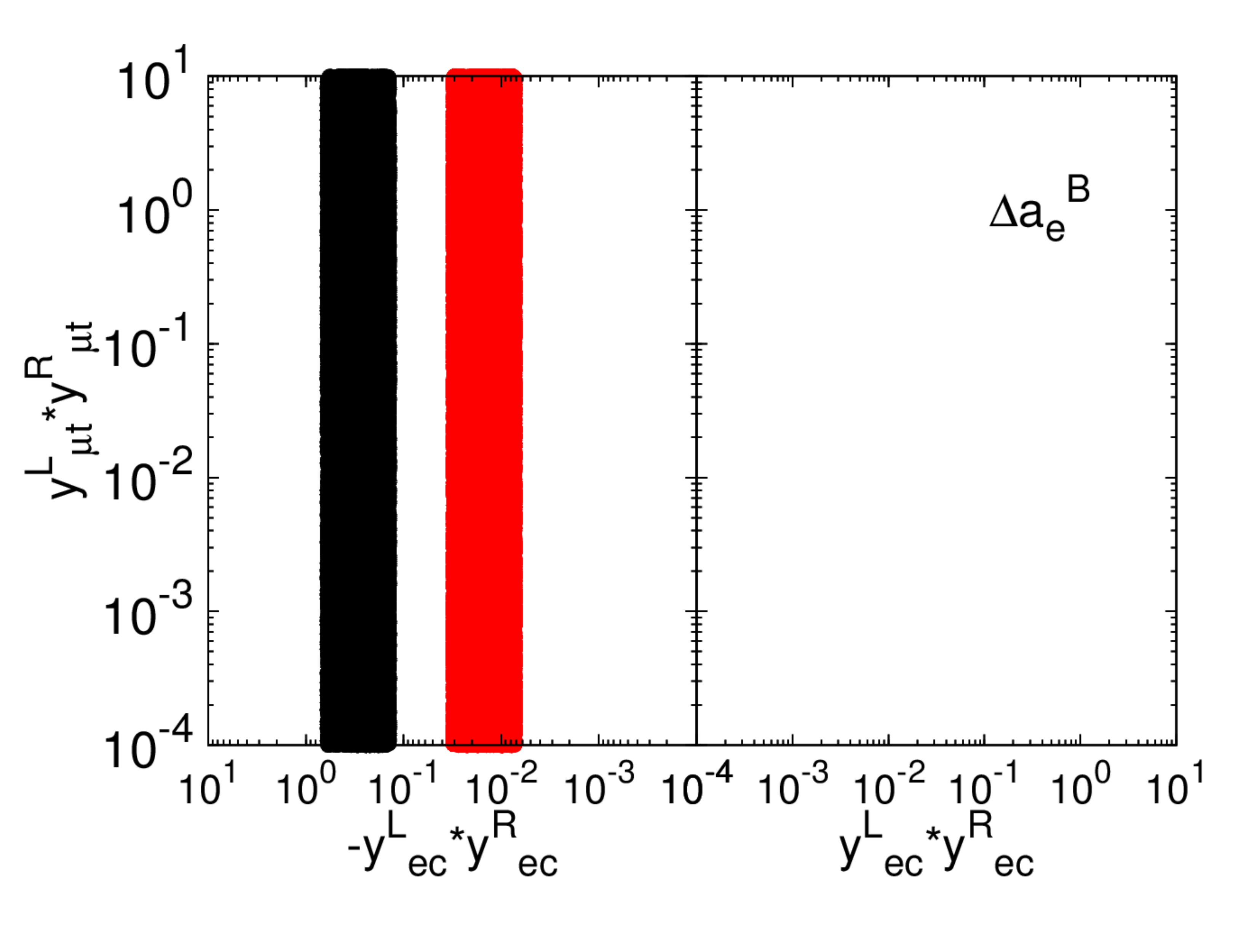}
\includegraphics[height=1.5in,angle=0]{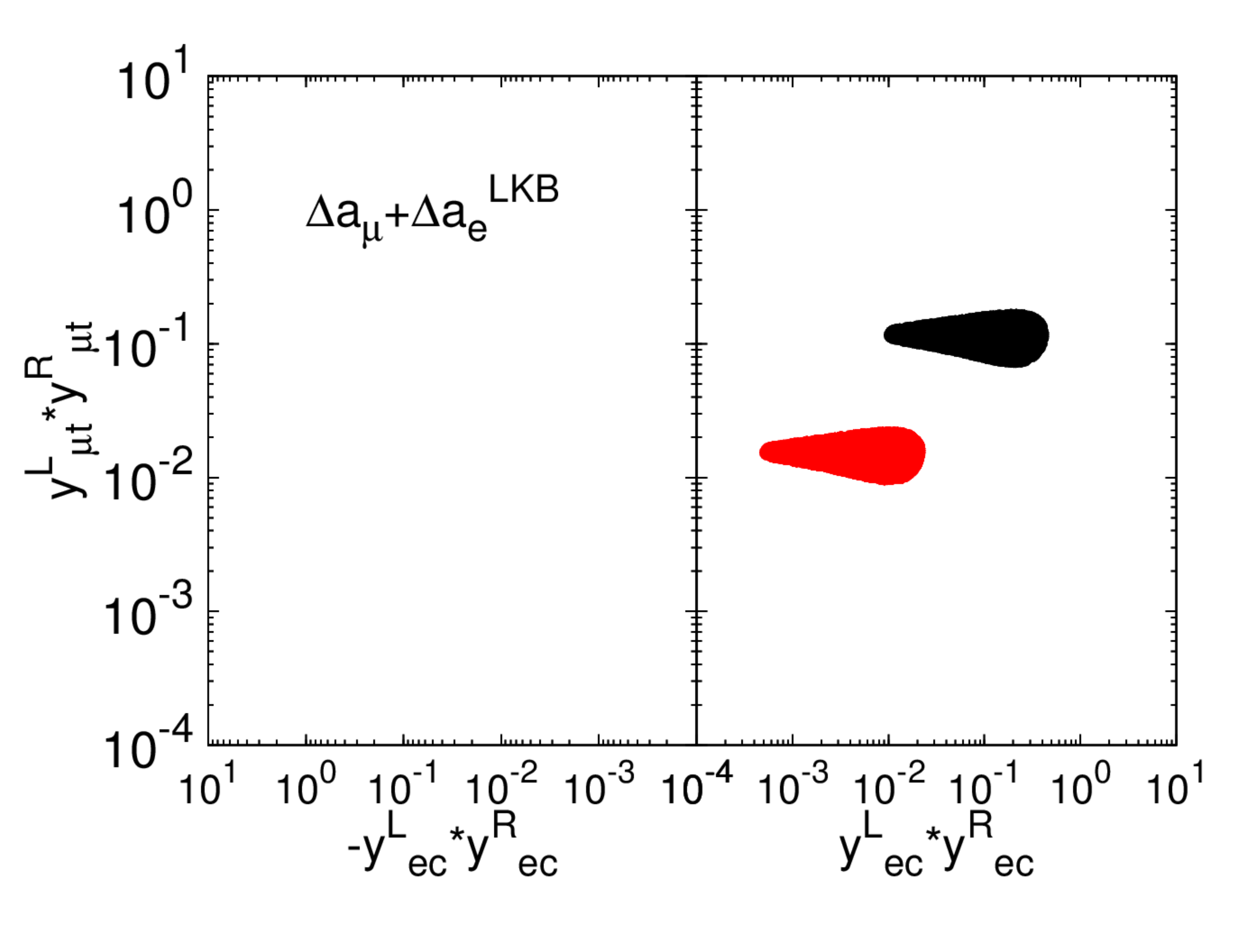}
\includegraphics[height=1.5in,angle=0]{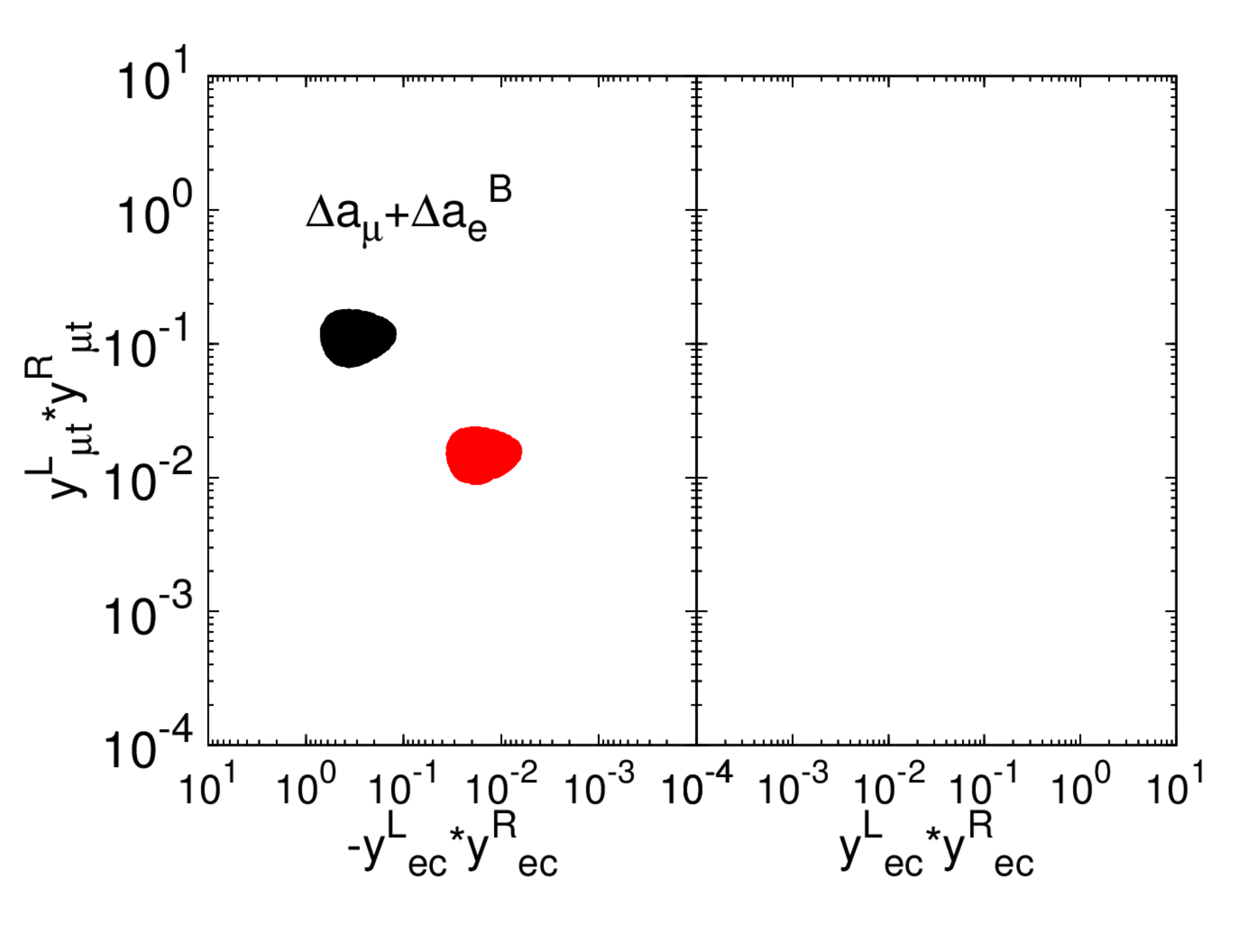}
\caption{\small \label{fig:S1_LQ}
{\bf $S_1$-LQ:} The 1$\sigma$ regions preferred by $\Delta a_\mu$, $\Delta a^{\rm LKB}_e$, and $\Delta a^{\rm B}_e$ (upper panels), and 
 from a combined fit to $\Delta a_\mu$ and $\Delta a^{\rm LKB}_e$, and
 $\Delta a_\mu$ and $\Delta a^{\rm B}_e$ (lower panels), for $m_{S_1}=10$~TeV (black)
$m_{S_1}=2$~TeV (red). }
\end{figure}

\begin{figure}[t]
\centering
\includegraphics[height=1.5in,angle=0]{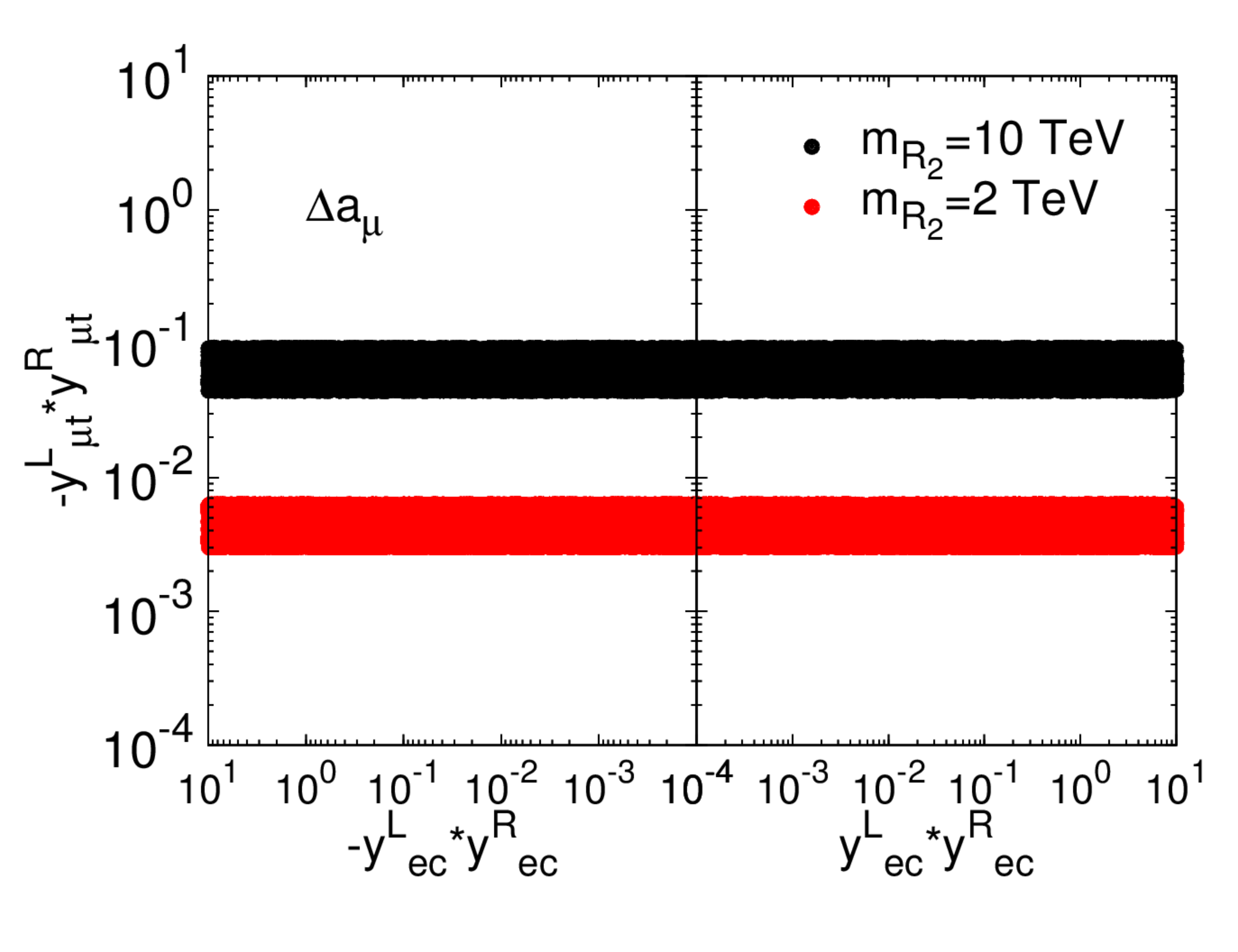}
\includegraphics[height=1.5in,angle=0]{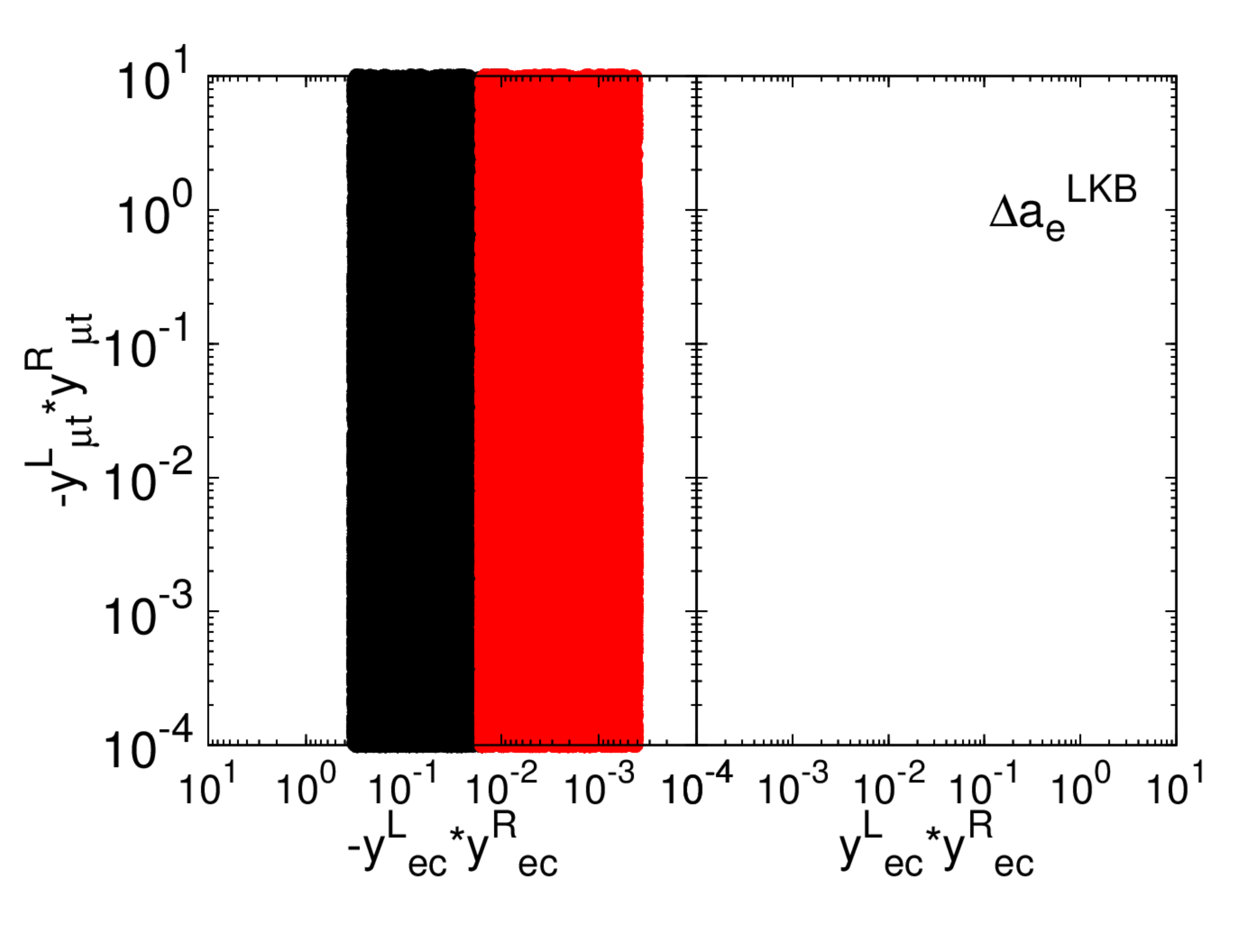}
\includegraphics[height=1.5in,angle=0]{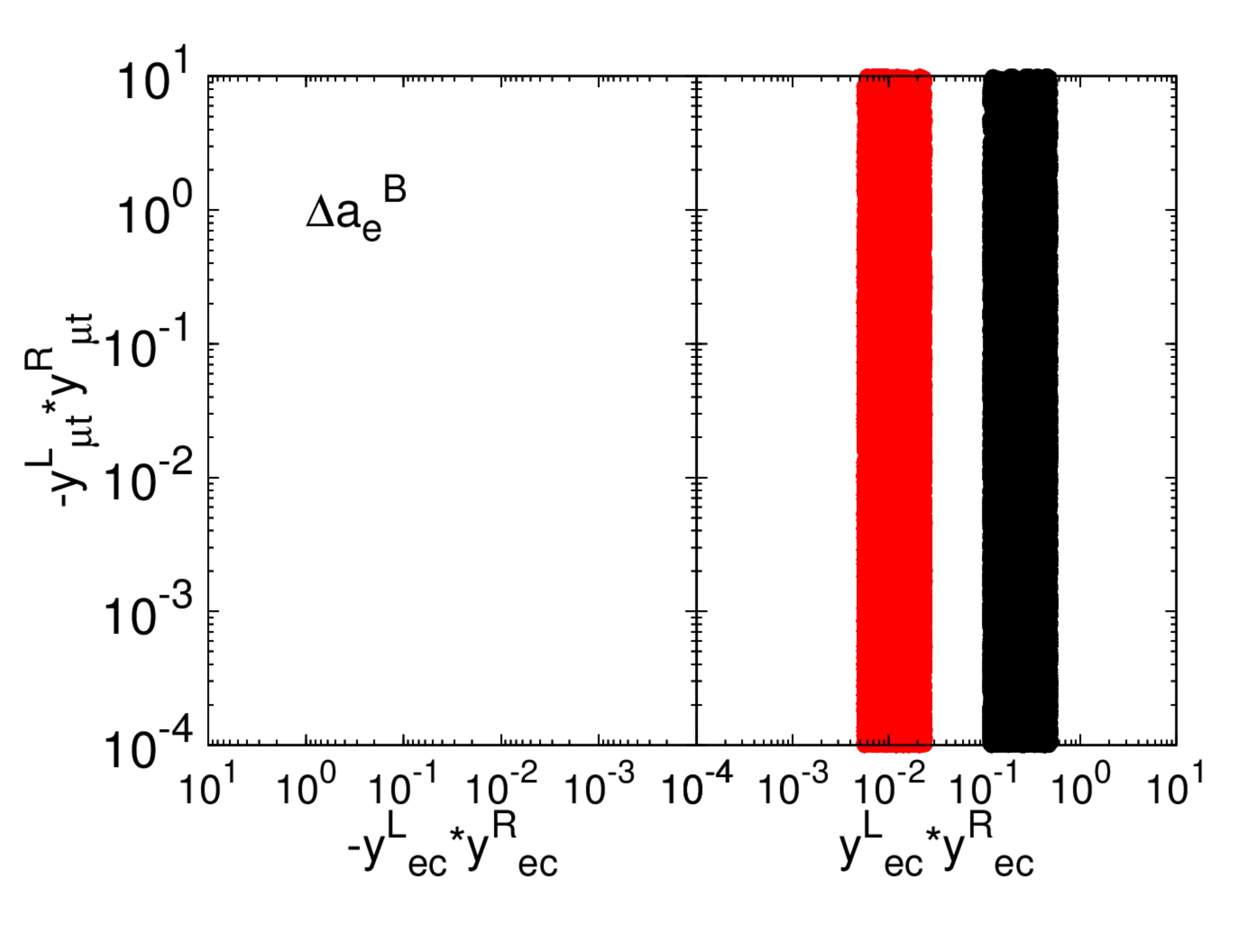}
\includegraphics[height=1.5in,angle=0]{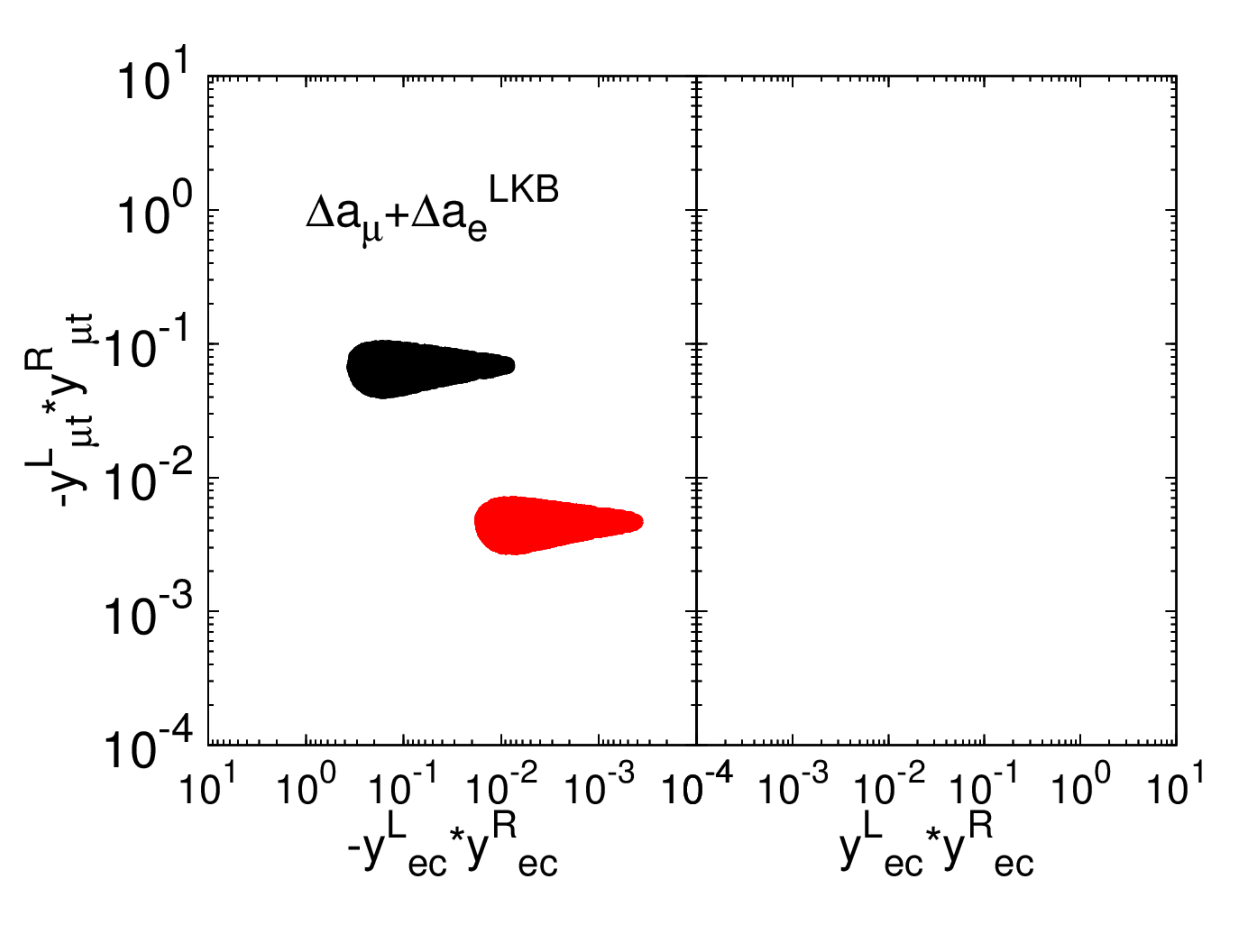}
\includegraphics[height=1.5in,angle=0]{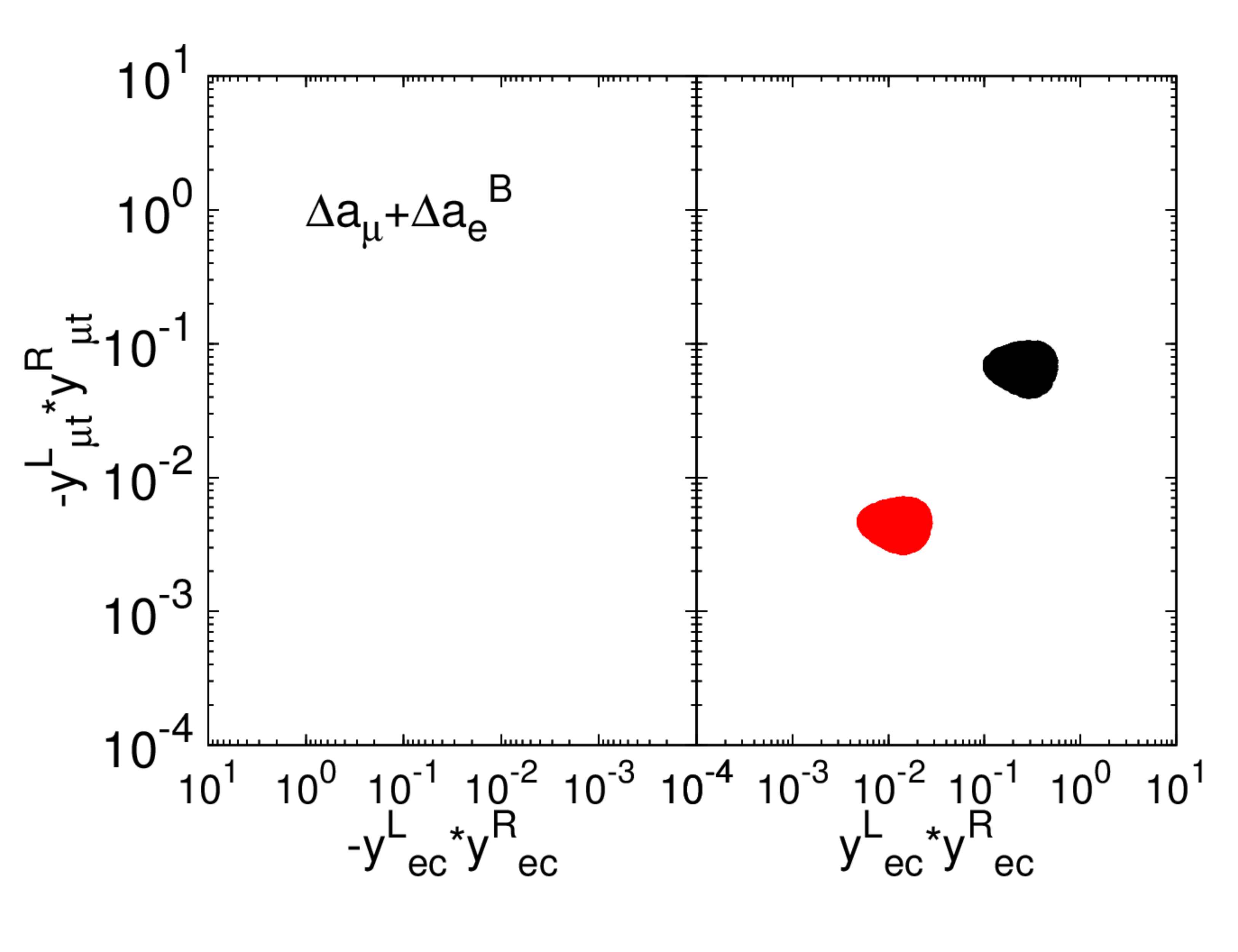}
\caption{\small \label{fig:R2_LQ}
{\bf $R_2$-LQ:} 
The 1$\sigma$ regions preferred by $\Delta a_\mu$, $\Delta a^{\rm LKB}_e$, and $\Delta a^{\rm B}_e$ (upper panels), and 
 from a combined fit to $\Delta a_\mu$ and $\Delta a^{\rm LKB}_e$, and
 $\Delta a_\mu$ and $\Delta a^{\rm B}_e$ (lower panels), for $m_{R_2}=10$~TeV (black)
$m_{R_2}=2$~TeV (red). 
}
\end{figure}

Taking the couplings to be real, we scan the parameter space
in two scenarios:
\begin{itemize}
\item {\bf $S_1$-LQ:} We fix $m_{S_1}=2$, 10 TeV, and vary $y^L_{ec}y^R_{ec}$ and $y^L_{\mu t}y^R_{\mu t}$. The results
are shown in Fig.~\ref{fig:S1_LQ} and Table~\ref{leptoq}.
\item {\bf $R_2$-LQ:} fixing $m_{R_2}=2$, 10 TeV, and varying $y^L_{ec}y^R_{ec}$, and $y^L_{\mu t}y^R_{\mu t}$. The results are
shown in Fig.~\ref{fig:R2_LQ} and Table~\ref{leptoq}.
\end{itemize}

\begin{table}[t]
\caption{\small \label{tab:min_LQ}
Best-fit points  for the scalar and doublet leptoquarks. The minimum $\chi^2$ value in all cases is 0.
}
\begin{adjustbox}{width=\textwidth}
\begin{tabular}{c|cccc|cccc}
\hline
\hline
    & \multicolumn{4}{c}{\bf $S_1$-LQ}  & \multicolumn{4}{c}{\bf $R_2$-LQ}   \\
\hline     
            & LKB & LKB & B & B & LKB  & LKB & B & B  \\
\hline
$m_{S_1(R_2)}/{\rm TeV}$     
            & 2 & 10 & 2 & 10 & 2  & 10 & 2 & 10  \\
$y^L_{ec}y^R_{ec}$     
            & $1.1\times 10^{-2}$ & 0.20 & $-1.9\times 10^{-2}$ & -0.37 & $-8.3\times 10^{-3}$ & -0.16 & $1.5\times 10^{-2}$ & 0.30 \\
$y^L_{\mu t}y^R_{\mu t}$     
            & $1.6\times 10^{-2}$ & 0.12 & $1.5\times 10^{-2}$ & 0.12 & $-4.7\times 10^{-3}$ & $-6.9\times 10^{-2}$ & $-4.7\times 10^{-3}$ & $-6.8\times 10^{-2}$  \\
\hline
\hline
\end{tabular}
\end{adjustbox}
\label{leptoq}
\end{table}

\section{Summary}

In light of the recent measurement of the muon anomalous magnetic moment by the Muon g-2 experiment, 
we examine three model frameworks as explanations
of the $(g-2)_{e,\mu}$ discrepancy with standard model expectations.   
We considered
i) axion-like particles with masses $\lesssim\mathcal{O}(1)$ GeV 
and couplings to charged leptons and photons, which yields a contribution to the 2-loop light-by-light diagram for $(g-2)_{e,\mu}$.
ii) Two-Higgs-doublet models with four Yukawa structures:
Type-I, II, X (lepton-specific), and Y (flipped),
where the CP-odd scalar with mass $\lesssim\mathcal{O}(100)$ GeV 
gives the main contribution to $(g-2)_{e,\mu}$ up to 2-loop Barr-Zee diagrams.
iii) Scalar leptoquarks, $S_1\sim ({\bf 3},{\bf 1},-1/3)$ 
and $R_2 \sim ({\bf 3},{\bf 2},7/6)$, 
where the Yukawa couplings are assigned 
as up-type mass-diagonal basis to avoid constraints from $\mu \to e \gamma$.
Then the mixed-chiral charm-electron and top-muon Yukawa couplings contribute to $(g-2)_e$
and $(g-2)_\mu$, respectively.

We find that accounting for other constraints, all scenarios except the Type-I, Type-II and Type-Y two-Higgs-doublet models
easily accommodate the data.

\section*{Acknowledgements}  
D.M. is supported in
part by the U.S. DOE under Grant No. de-sc0010504.
P.T is supported by National Research Foundation of Korea (NRF-2020R1I1A1A01066413).


\end{document}